\documentclass[12pt]{article}

\usepackage{scicite}
\usepackage{times}
\usepackage{url}
\usepackage{graphicx}
\usepackage{xcolor} 
\usepackage{hyperref}

\newenvironment{sciabstract}{%
\begin{quote} \bf}
{\end{quote}}

\def\aap{Astron. \& Astrophys.}                
\newcommand{\aapr}{Astron. Astrophys. Rev.}
\newcommand{\aj}{Astron. J.}
\newcommand{\apjs}{Astrophys. J.}
\newcommand{\apj}{Astrophys. J.}
\newcommand{\apjl}{Astrophys. J.}
\newcommand{\mnras}{Mon. Not. R. Astro. Soc.}
\newcommand{\pasa}{Publ. Astron. Soc. Australia}
\newcommand{\nat}{Nature}

\newcommand{\pasp}{Pub. Astron. Soc. Pacific}

\newcommand{\pasj}{Publ. Astron. Soc. J.}
\newcommand{\pccm}{\ensuremath{\mathrm{pc~cm}^{-3}}}

\newcommand{\arcsec}{$^{\prime\prime}$}
\newcommand{\arcmin}{$^{\prime}$}
\newcommand{\zfrb}{\ensuremath{1.016 \pm 0.002}}

\newcommand{\frbname}{FRB\,20220610A}
\newcommand{\rmfrb}{$215 \pm 2$\,rad\,m$^{-2}$}
\newcommand{\sigmarm}{$0.6$\,rad\,m$^{-2}$}
 
\newcommand{\taufrb}{$0.511\pm0.012$\,ms}

\newcommand{\dminit}{\ensuremath{1457.6\,\mathrm{pc\,cm}^{-3}}}
\newcommand{\dmfrb}{\ensuremath{1458.15_{-0.55}^{+0.25}\,\mathrm{pc\,cm}^{-3}}}

\newcommand{\dmcosmic}{\ensuremath{{\rm DM}_{\rm cosmic}}}

\newcommand{\dmhost}{\ensuremath{{\rm DM}_{\rm host}}}
\newcommand{\dmmw}{\ensuremath{{\rm DM}_{\rm MW}}}
\newcommand{\dmhalo}{\ensuremath{{\rm DM}_{\rm halo}}}
\newcommand{\Hnot}{\ensuremath{{H_{0}}}}
\newcommand{\hubbunit}{\ensuremath{\rm km \, s^{-1} \, Mpc^{-1}}}
\newcommand{\omegabhh}{\ensuremath{\Omega_b h^2}}

\newcommand{\emax}{\ensuremath{E_{\rm max}}}
\newcommand{\lemaxerg}{\ensuremath{\log_{10} (E_{\rm max}/{\rm erg})}}

\title{A luminous fast radio burst that probes the Universe at redshift $1$}

\author
{S.~D.~Ryder$^{1,2}$, K.~W.~Bannister$^{3}$, S.~Bhandari$^{4,5}$, A.~T.~Deller$^{6}$, R.~D.~Ekers$^{3,7}$,\\ 
M.~Glowacki$^{7}$, A.~C.~Gordon$^{8}$, K.~Gourdji$^{6}$, C.~W.~James$^{7}$,\\C.~D.~Kilpatrick$^{8,9}$, W.~Lu$^{10,11}$, L.~Marnoch$^{1,2,3,12}$, V.~A. Moss$^{3}$,\\  J.~X. Prochaska$^{13,14}$, H.~Qiu$^{15}$, E.~M.~Sadler$^{16,3}$, S.~Simha$^{13}$,\\ M.~W.~Sammons$^{7}$, D.~R.~Scott$^{7}$, N.~Tejos$^{17}$, R.~M.~Shannon$^{6,\ast}$
\\
\footnotesize{$^{1}$School of Mathematical and Physical Sciences, Macquarie University, NSW 2109, Australia}\\
\footnotesize{$^{2}$Astrophysics and Space Technologies Research Centre, Macquarie University, Sydney, NSW 2109, Australia}\\
\footnotesize{$^{3}$Australia Telescope National Facility, Commonwealth Science and Industrial Research Organisation (CSIRO),}\\
\footnotesize{ Space and Astronomy, PO Box 76, Epping, NSW 1710, Australia}\\
\footnotesize{$^{4}$The Netherlands Institute for Radio Astronomy (ASTRON), 7991 PD Dwingeloo, The Netherlands}\\
\footnotesize{$^{5}$Joint institute for Very Long Baseline Interferometry in Europe,   7991 PD Dwingeloo, The Netherlands}\\
\footnotesize{$^{6}$Centre for Astrophysics and Supercomputing, Swinburne University of Technology, Hawthorn,VIC 3122, Australia}\\
\footnotesize{$^{7}$International Centre for Radio Astronomy Research, Curtin Institute of Radio Astronomy,}\\
\footnotesize{Curtin University, Perth, Western Australia, Australia.}\\
\footnotesize{$^8$Center for Interdisciplinary Exploration and Research in Astrophysics,  Northwestern University,}\\
\footnotesize{Evanston, IL 60208, USA} \\
\footnotesize{$^{9}$  Department of Physics and Astronomy, Northwestern University, Evanston, IL 60208, USA} \\
\footnotesize{$^{10}$Department of Astronomy  University of California, Berkeley, CA 94720, USA} \\
\footnotesize{$^{11}$ Theoretical Astrophysics Center, University of California, Berkeley, CA 94720, USA} \\
\footnotesize{$^{12}$ Australian Research Council Centre of Excellence for All-Sky Astrophysics in 3 Dimensions (ASTRO 3D), Australia} \\
\footnotesize{$^{13}$ Department of Astronomy and Astrophysics, University of California,}
\footnotesize{Santa Cruz, CA 95064, USA}\\
\footnotesize{$^{14}$  Kavli Institute for the Physics and Mathematics of the Universe, Kashiwa, 277-8583, Japan}\\
\footnotesize{$^{15}$SKA Observatory, Jodrell Bank, Lower Withington, Macclesfield, SK11 9FT, UK}\\
\footnotesize{$^{16}$Sydney Institute for Astronomy, School of Physics, University of Sydney, NSW 2006, Australia}\\
\footnotesize{$^{17}$Instituto de F\'isica, Pontificia Universidad Cat\'olica de Valpara\'iso, Casilla 4059, Valpara\'iso, Chile}\\
\footnotesize{$^\ast$Corresponding author: E-mail: rshannon@swin.edu.au.}
}

\date{}
\usepackage{amsmath}
\usepackage{amssymb}

\begin{document} 


\baselineskip24pt


\maketitle 
\clearpage


\begin{sciabstract}
Fast radio bursts (FRBs) are millisecond-duration pulses of radio emission originating from extragalactic distances. 
Radio dispersion on each burst  is imparted by intervening  plasma mostly located in the intergalactic medium.
We observe a burst, \frbname, in a morphologically complex host galaxy system at redshift $\zfrb$.
The burst redshift and dispersion 
are consistent with passage through a substantial column of material from the intergalactic medium. 
The burst shows evidence for passage through additional turbulent magnetized  plasma, potentially associated with the host galaxy. We use the burst energy of $2 \times 10^{42}$\, erg, to revise the maximum energy of an FRB.
\end{sciabstract}

Fast radio bursts  (FRBs, \cite{2007Sci...318..777L,2022A&ARv..30....2P}) are transient radio sources that last a few milliseconds emitted by extragalactic sources.
Free electrons along the  path between the FRB source and the Earth impart 
a frequency-dependent time delay (dispersion) on the radio signal. This dispersion can be used to  measure the column density of free electrons (quantified by the dispersion measure, DM) between the FRB source and observer. FRBs localized to host galaxies at different redshifts exhibit a positive correlation between extragalactic DM and host redshift, known as the Macquart relation \cite{2020Natur.581..391M}. 
This relation has been used to measure the cosmic baryon fraction and the expansion rate of the Universe \cite{2022MNRAS.516.4862J}. This relation has been measured using identified FRB host galaxies at relatively low redshifts, $z \lesssim 0.5$. Some unlocalized FRBs (with unknown host galaxies) have DMs consistent with  $z > 1$ \cite{2018MNRAS.475.1427B}; however an FRB associated with a galaxy at $z=0.241$ had a high DM that would have implied $z>1$ \cite{2022Natur.606..873N}. 
This indicates that estimates of redshift from DM alone can be misled by plasma within the host galaxy, which
also imparts a contribution to the DM, in addition to that of the intergalactic medium.

\pagebreak
\noindent{\bf Observations of  FRB 20220610A}

FRBs have been searched for \cite{2018Natur.562..386S} and localized \cite{2019Sci...365..565B}  using the Australian Square Kilometre Array Pathfinder (ASKAP, \cite{2021PASA...38....9H}),  a radio interferometer in Western Australia  comprising 36, 12-m antennas. 
Each antenna is equipped with phased-array receiving systems, which provide 36 beams across the focal plane, covering approximately 30 square degrees.
FRBs are searched for in real time using the incoherent sum of the intensities of each antenna in each beam.  
When an FRB is detected, voltage buffers are downloaded, correlated, calibrated, and imaged, enabling the position of the burst to be measured to an absolute precision of typically a few tenths of an arcsecond \cite{2021PASA...38...50D,online}.

We detected \frbname~in observations around the previously known burst FRB~20220501C, but the two bursts are not related \cite{online}. The observations were centred at a frequency of $1271.5$\,MHz and had a time resolution of 1.18\,ms.  The dispersion of \frbname\ indicated a DM of \dmfrb.  This is higher than all but one of the $55$ FRBs previously observed using ASKAP \cite{2022MNRAS.516.4862J}.   
A dedispersed dynamic spectrum of the burst is shown in Figure~1, and properties of the burst are listed in Table~1. The burst does not show the 10-100 MHz modulation in the spectrum characteristic of many lower DM, high Galactic latitude ASKAP-detected FRBs \cite{2018Natur.562..386S}.

The 2\,s dispersive sweep of the burst across the instrument bandwidth and 2.4\,s latency in the detection system resulted in only the lowest $88$\,MHz of the burst being captured in the $3.1$\,s-duration voltage buffer.
This was sufficient to localize the burst to a precision of 0.5 arcsec. 
We  used the voltage data to reconstruct the high time resolution and polarimetric properties of the burst \cite{online}.
After correcting for dispersive smearing, the burst shows an exponentially decreasing tail (Fig.~1D), which is consistent with scatter broadening due to turbulence in intervening plasma \cite{1968Natur.218..920S}.  We measure the pulse broadening time  to be \taufrb\ at a reference frequency of $1147.5$\,MHz,
assuming a $\nu^{-4}$ frequency dependence.

Ordered magnetic fields in astrophysical plasmas add additional, polarization-dependent dispersion. This manifests as wavelength-dependent variation in the linear polarization position angle, referred to as Faraday rotation \cite{2022arXiv220915113M}.
The burst exhibits Faraday rotation, with a rotation measure (RM) of \rmfrb. 
After correcting for this Faraday rotation,  we find the burst had a linear polarization fraction of $ 96 \pm 1$\,\%. 
The high fractional linear polarization allows us to place a 67\% upper limit on the Faraday dispersion ($\sigma_{\rm RM}$ $<$  \sigmarm)  induced by fluctuations in rotation measure in intervening turbulent plasma.
Higher levels of Faraday dispersion have been detected for other FRBs \cite{2022Sci...375.1266F}.
The burst also shows modest fractional circular polarization of $10\pm 1\%$.  While instrumental artifacts can induce spurious circular polarization, we do not see any correlation between Stokes polarization parameters $U$ and $V$ in the spectrum, which would be expected for an instrumental effect \cite{online}. The FRB was located approximately $4$\,arcmin from the beam center, which makes off-axis leakage effects less likely \cite{2020PASA...37...48M}.  Circular polarization has been observed in some FRBs and could either be intrinsic to the burst  \cite{Cho2020}  or  result from propagation through relativistic plasma in the immediate source environment \cite{2022MNRAS.512.3400K}.

\vskip 5mm
\noindent{\bf Host-galaxy properties}

We performed follow-up ground-based optical and infrared observations with the Very Large Telescope (VLT) and the W.~M. Keck Observatory to identify and characterize the host galaxy of \frbname \cite{online}.
The images (Fig. 2A-C) show an object coincident with the source that has an extended, multi-component morphology. 
We label the optical source that overlaps the radio position of the FRB as component (a), and two adjacent sources as components (b) and (c) (Fig. 2A). 
We use a Bayesian method to assess the chance of coincidence between transients and host galaxies \cite{2021ApJ...911...95A}, finding greater than $ 99.99\%$ confidence that the FRB is associated with component (a).

We performed  broad band optical and infrared spectroscopy of components (a), (b) and (c) (Fig.~2D-E) \cite{online}. We identify two emission lines in the spectra as the [O\,{\sc ii}] 3726 and 3729~\AA\ doublet, most prominently in component (b) (Fig. S2).  From this we measure the redshift of each component, finding they are all consistent with $z=$ \zfrb. 

We estimate the total mass of the three components combined  to be $10^{10}$ solar masses, with a star formation rate of $\sim 0.42$ solar masses per year \cite{online}.
These values, in addition to the host metallicity and star formation history are consistent with those of nearby FRB hosts \cite{Bhandari2020,Bhandari2022}, but the source morphology is markedly different.
Properties of the host galaxy are listed in Table 1.

The presence of two bright components (a) and (c) separated by 2.0~arcseconds (which corresponds to a distance of 16\,kpc at that redshift), and the diffuse feature (b) between them, is consistent with two galaxies interacting or merging, or a compact galaxy group.  
It is also possible that the morphology is due to internal structure within a single galaxy; at these redshifts about half of all galaxies have clumpy morphologies \cite{Guo2015}. We regard the latter possibility as unlikely, due to the large spatial separation between the components.
Only component (a) is detected in the near-infrared ($K_{\rm s}$-band) image (Fig. 2C), indicating it hosts an older stellar population than the other components. Component (a) is also displaced from the centroid of the total optical light in $g$- and $R$-bands,  contrary to what would be expected if it was the nuclear bulge of a single  galaxy.

\vskip 5mm
\noindent{\bf Extending the Macquart Relation}

We used the measured properties of \frbname\ to probe the Macquart relation to $z\sim 1$, by comparing with predictions for its DM based on previous fits to the relation at $z \le 0.522$.
Figure 3 shows the relationship between DM and redshift for the FRBs detected by ASKAP \cite{2022MNRAS.516.4862J}.
We restrict our analysis to the ASKAP sample \cite{2022MNRAS.516.4862J} to minimize observing-system-dependent selection effects.
We do not re-fit the Macquart relation. Doing so in an unbiased way would require analyzing the entire updated FRB sample from ASKAP.

After subtracting a  model  Milky Way foreground contribution to the observed DM
\cite{online},
we estimate the non-Galactic DM of \frbname\ to be $\approx 1376 \, \rm pc \, cm^{-3}$,
indicating a high column density of ionized gas between the FRB and Earth. This is higher than the DM expected from the Macquart relation, by approximately $650$\,pc\,cm$^{-3}$, and is a $2.4\sigma$  excess \cite{online}. 
If the excess is real and originates from the host galaxy, the implied electron column density is $1300_{-320}^{+170}$\,pc\,cm$^{-3}$ in the host rest frame \cite{online}, with the uncertainty reflecting the intrinsic variation in the contribution from the intergalactic medium.

\vskip 5mm
\noindent{\bf Interpretation of the Dispersion Measure}

We use the scatter broadening and Faraday rotation of \frbname\ to investigate the properties of plasma at  $z\approx1$ .
The possible  dispersion excess could arise from 
any combination of gas in the immediate vicinity of the source, the interstellar medium of the host galaxy, or foreground  gas along the line of sight,  each of which could potentially harbor turbulent  magnetized plasma. 
The absence of strong ($\gtrsim 10^4$ rad\,m$^{-2}$) Faraday rotation  or detectable depolarization is unlike the repeating FRB\,20190520B, which also shows excess dispersion \cite{2022Natur.606..873N,2022Sci...375.1266F}. We suggest the excess dispersion for \frbname\ originates in a less magnetoionically active plasma than FRB\,20190520B, such as in the interstellar medium of the host galaxy, rather than  in the circumburst media hypothesized for other sources \cite{online}.   
Models of galaxy interstellar media imply that the DM of a typical spiral galaxy is unlikely to exceed a few hundred \pccm\ except in  edge-on systems \cite{2015RAA....15.1629X,2023MNRAS.518..539M}. Higher DM values can plausibly be produced by high-density clumps of gas within the host galaxy, particularly  at $z\sim1$ where galaxies have a substantially higher fraction of their baryons in gas (rather than stars) compared to $z\sim 0$ \cite{2022ApJ...941L...6C}.
Alternatively the dispersion could originate from structure in the foreground intergalactic medium, or additional ionized material associated with the possible galaxy merger between components (a), (b), and (c).

Our DM analysis confirms inferences from other techniques 
\cite{kim2021} that the gas of the intergalactic medium is highly ionized.
The detection of an FRB at $z>1$ allows us
to study the ionized plasma towards, around, and within the host galaxy. We expect a sightline to $z=1$ to intersect
the halos of several galaxies similar in mass and size to the Milky Way \cite{ProchaskaZheng2019}, yielding further insight into their properties \cite{20181112}.

\vskip 5mm
\noindent{\bf Interpretation of the burst energy}

The measured bandwidth-averaged fluence of \frbname\ is $45~\pm~5$\,Jy\,ms (1 Jy = 10$^{-23}$\,erg\,s$^{-1}$\,cm$^{-2}$\,Hz$^{-1}$), implying an isotropic-equivalent spectral energy density of $(6.4~\pm~0.7) \times 10^{32}$\,erg\,Hz$^{-1}$ and a burst energy of $(6.4~\pm~0.7)\times 10^{41}$\,erg \cite{online}.  We derived the burst energy by assuming an intrinsic 1\,GHz bandwidth,  and did not apply a redshift-dependent correction to the burst spectral energy distribution (i.e., K-correction) \cite{online}.  This value exceeds the characteristic maximum energy \emax\ derived by previous FRB population models by a factor of 3.5 \cite{James2022Meth,online}. 
It is unknown whether FRBs are emitted isotropically or only over a limited beaming angle, which would affect the inferred energetics and could vary between FRB sources or repeat FRBs from single sources.
Assuming isotropic emission, we have re-fitted the FRB burst energy distribution by adding \frbname\ to the sample of FRBs used by \cite{James2022Meth}, finding the best-fit \emax\ to increase by a factor of $2.7$, to $10^{41.7~\pm~ 0.2}$\,erg (i.e., the maximum energy density becomes $\log_{10} E_\nu\,[{\rm erg\,Hz^{-1}}] = 32.7 \pm 0.2$).
In Figure 4, we  contrast the burst to the fluence of the brightest radio pulse observed from a galactic magnetar, which is a factor of $10^5$ less luminous than the burst observed from \frbname, and the wider sample of FRBs.

If a K-correction is applied, assuming a  spectral index for FRB emission similar to that of the burst population found by ASKAP \cite{2019ApJ...872L..19M,online},  we find the burst energy integrated over the instrument bandwidth is $\sim 2\times10^{42}\rm\,erg$, which is higher than most localized FRBs (Figure 4) 
\cite{2019Sci...365..565B,20181112,2020Natur.581..391M,2022MNRAS.516.4862J,2019Natur.572..352R,2020Natur.586..693L,2022Natur.606..873N,2019MNRAS.486.3636P,2022MNRAS.512.3400K,2021ATel14556....1H, 2019ApJ...876L..23H,2020Natur.577..190M,2023MNRAS.520.2281N}. 
This constrains emission models of FRBs, because the electric field strength at the source can be estimated independently of the beaming angle. The calculation assumes no amplification by gravitational or plasma lensing. In one class of models, FRB emission is produced near the surface of a neutron star. From the luminosity of the burst $\sim 3\times10^{46}\rm\, erg\,s^{-1}$ in the host galaxy's rest frame, we infer an electric field strength of 
$4.2\times10^{12} (r/10\mathrm{\,km})^{-1}$\,V\,m$^{-1}$, for a linearly polarized wave, where  $r$ ($\sim$10\,km) is the curvature radius of the neutron star's magnetic field. 
This is a few percent of the Schwinger critical field strength at a neutron star surface, in which an electric field aligned parallel to the local magnetic field, would be quickly screened by copious electron-position pair production  \cite{2019MNRAS.483L..93L}. 
This would suppress the FRB rate above the Schwinger luminosity of approximately $2 \times 10^{47}$\,erg\,s$^{-1}$ \cite{2019MNRAS.483L..93L}. 
In another class of models, FRBs are produced in a shock driven by relativistic ejecta associated with the flare of a highly magnetized neutron star interacting with the neutron star wind. In these models, the radiative efficiency in the shock is very low ($\lesssim 10^{-5}$) and hence the required  energy in the ejecta would be $\gtrsim 10^{47}\rm\, erg$, with the total flare energy being even higher. \frbname\ and other high luminosity FRBs challenge both model classes.


\noindent
{\bf Acknowledgements:} We thank K. Heintz for contributions to the FURBY project, and N. F\"{o}rster-Schreiber and K. Glazebrook for discussions.
We thank ATNF staff for supporting the CRAFT observations with ASKAP. 
Based on observations collected at the European Southern Observatory under ESO programmes 0105.A-0687 and 1108.A-0027.
This scientific work uses data obtained from Inyarrimanha Ilgari Bundara / the Murchison Radio-astronomy Observatory. We acknowledge the Wajarri Yamaji People as the Traditional Owners and native title holders of the Observatory site. CSIRO’s ASKAP radio telescope is part of the Australia Telescope National Facility (https://ror.org/05qajvd42). Operation of ASKAP is funded by the Australian Government with support from the National Collaborative Research Infrastructure Strategy. ASKAP uses the resources of the Pawsey Supercomputing Research Centre. Establishment of ASKAP, Inyarrimanha Ilgari Bundara, the CSIRO Murchison Radio-astronomy Observatory, and the Pawsey Supercomputing Research Centre are initiatives of the Australian Government, with support from the Government of Western Australia and the Science and Industry Endowment Fund.
Some of the data presented herein were obtained at the W. M. Keck Observatory, which is operated as a scientific partnership among the California Institute of Technology, the University of California, and the National Aeronautics and Space Administration. The Observatory was made possible by the generous financial support of the W. M. Keck Foundation. We recognize and acknowledge the very important cultural role and reverence that the summit of Maunakea has always had within the indigenous Hawaiian community.  We are most fortunate to have the opportunity to conduct observations from this mountain.
{\bf Funding:}
 SB is supported by a Dutch Research Council (NWO) Veni Fellowship (VI.Veni.212.058).
 ATD and KG acknowledge support  through Australian Research Council (ARC) Discovery Project (DP) DP200102243.
 ATD and RMS acknowledge support through Australian Research Council ARC DP DP220102305.
CWJ and MG acknowledge support through ARC DP DP210102103.
 RMS acknowledges support through ARC Future Fellowship FT190100155. 
 JXP, SS, CK, ACG, and NT acknowledge support from National Science Foundation grants AST-1911140, AST-1910471, and AST-2206490.
JXP is a Simons Pivot Fellow.
NT acknowledges support through FONDECYT grant 11191217.
{\bf Author contributions:}
SDR, ACG, LM, NT, JXP, and SS collected and analyzed the optical and infrared imaging and spectroscopic observations. KWB developed the FRB search and localization systems with contributions from SB and RMS.  ATD, MG, KG, and DRS performed the FRB localization and high time resolution processing. RDE and EMS provided comments on the manuscript and the interpretation of the host-galaxy properties.
CWJ and WL interpreted the burst energetics and implications for the FRB emission mechanism. VAM developed the observatory control systems that enabled FRB searches and calibration observations. HQ and MS measured the burst temporal properties. SDR, KWB, ATD, CWJ, and JXP contributed to the drafting of the manuscript. RMS led the drafting of the manuscript, coordinated the FRB searches, and measured  and interpreted the burst polarimetric properties. 
{\bf Competing Interests:} The authors declare no competing interests.  {\bf Data and materials availability:} The raw ASKAP observations, reduced radio images and burst data, reduced optical/infrared images and reduced optical and infrared spectra are available from the  Swinburne Data Portal \cite{data}.  The raw VLT observations are available from the \href{http://archive.eso.org/eso/eso\_archive\_main.html}
{ESO archive} (http://archive.eso.org/eso/eso\_archive\_main.html) under programme IDs 0105.A-0687 and 1108.A-0027. \\

\noindent
{\bf Supplementary Materials:}\\
Materials and Methods\\
Supplementary Text\\
Fig S1 $-$ S8\\
Tables S1 $-$ S2 \\
References (43 $-$ 108) \\

\clearpage
\begin{table}[]
{\bf Table 1}.  {\bf Properties of \frbname\ and its host galaxy.} The fluence was derived from filterbank data that includes the full ASKAP bandwidth.   UTC is the Coordinated Universal Time standard.  S/N is the signal to noise ratio. FWHM is full width at half maximum.  The brightness of the host is expressed in magnitudes (mag) on the AB system. The star formation rate (SFR) is in solar masses per year ($M_\odot$\,yr$^{-1}$).  The integrated star formation in the last 100 Myr is listed in the row 100 Myr SFR.  The metallicity of the host galaxy, $Z$, expressed as a (logarithmic) fraction of Solar metallicity $Z_\odot$ is listed in the row log(Metallicity).
\vskip 5mm
\begin{tabular}{l  r}
\hline
\multicolumn{2}{c}{Measured burst properties} \\
\hline
Dispersion Measure (DM)  &  \dmfrb\\
Topocentric arrival time at 1104~MHz (UTC) & 2022-06-10  22:26:44.313  \\
Fluence  & $45 \pm 5$ Jy\,ms \\
Right Ascension (J2000 equinox) & $23^{\rm h} 24^{\rm m} 17.569^{\rm s} \pm 0.040^{\rm s}$ \\
Declination (J2000 equinox) & $-33^{\circ} 30^{\prime} 49^{\prime\prime}.37 \pm 0^{\prime\prime}$.50 \\
Galactic Longitude  &  $8.83954^{\circ}$ \\
Galactic Latitude &  $-70.18569^{\circ}$  \\
Incoherent Detection S/N (1104 -- 1440 MHz) & 29.8 \\
Image S/N (1104-1152 MHz) & 81 \\
Rotation Measure & \rmfrb \\
\hline
\multicolumn{2}{c}{Inferred burst properties}\\
\hline
Intrinsic width (FWHM) & $0.41\pm 0.01$\,ms \\
Implied FRB isotropic energy density & $6 \times 10^{32}$~erg\,Hz$^{-1}$ \\
Milky-Way disk DM contribution & 31 pc\,cm$^{-3}$ \\
\hline
\multicolumn{2}{c}{Measured host galaxy properties} \\
\hline
Redshift   & \zfrb \\ 
\hline
\hline
\multicolumn{2}{c}{Photometry} \\
\hline
Band (Central wavelength, \AA) & Magnitude (mag)\\
\hline
$g$ (4700)           &  $24.15 \pm 0.07$ \\
$V$ (5510)          &  $23.89 \pm 0.13$ \\
$R$ (6580)   &  $23.78 \pm 0.06$ \\
$I$  (8060)         &  $22.17 \pm 0.07$ \\
$z$  (9620)         &  $21.95 \pm 0.12$ \\
$J$   (12200)          &  $21.97 \pm 0.07$ \\
$K_{\rm s}$  (21,460)   &  $22.08  \pm 0.12$ \\
\hline
\multicolumn{2}{c}{Inferred host galaxy properties} \\
\hline
Mass-weighted age & $1.02^{+1.64}_{-0.62}$ Gyr  \\
log(Stellar mass)  & $ 9.98^{+0.14}_{-0.07}$ M$_{\odot}$ \\
log(Total mass)  & $10.11^{+0.18}_{-0.07} $ M$_{\odot}$ \\
100 Myr SFR & $0.42^{+0.83}_{-0.37}$  M$_{\odot}$\,yr$^{-1}$ \\
log(Metallicity) & $-0.11^{+0.17}_{-1.68}$ $Z/Z_{\odot}$ \\
\hline
\label{tab:properties}
\end{tabular}
\end{table}

\clearpage

\begin{tabular}{cc}
    \centering
    \includegraphics[scale=0.6]{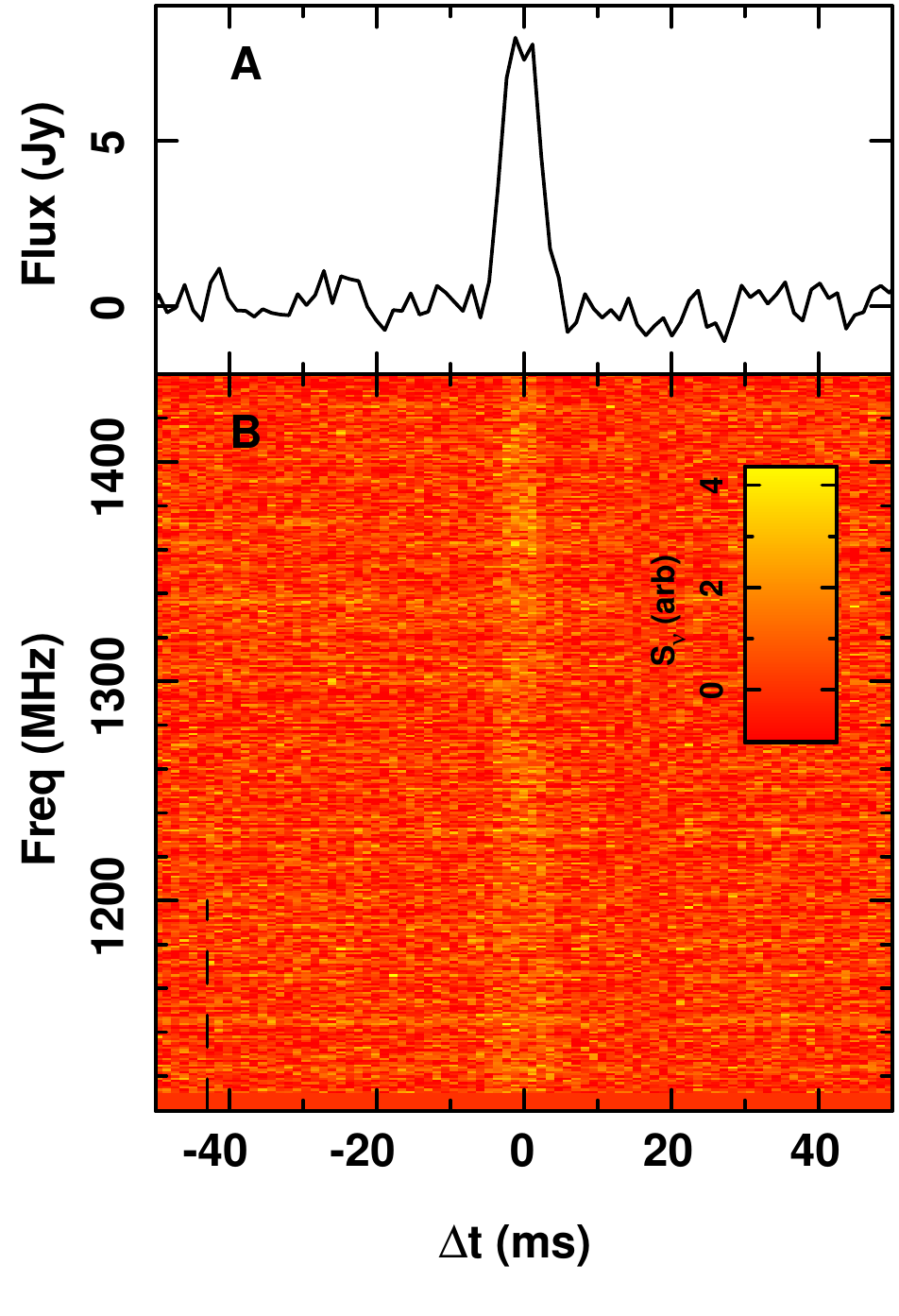} &  \includegraphics[scale=0.6]{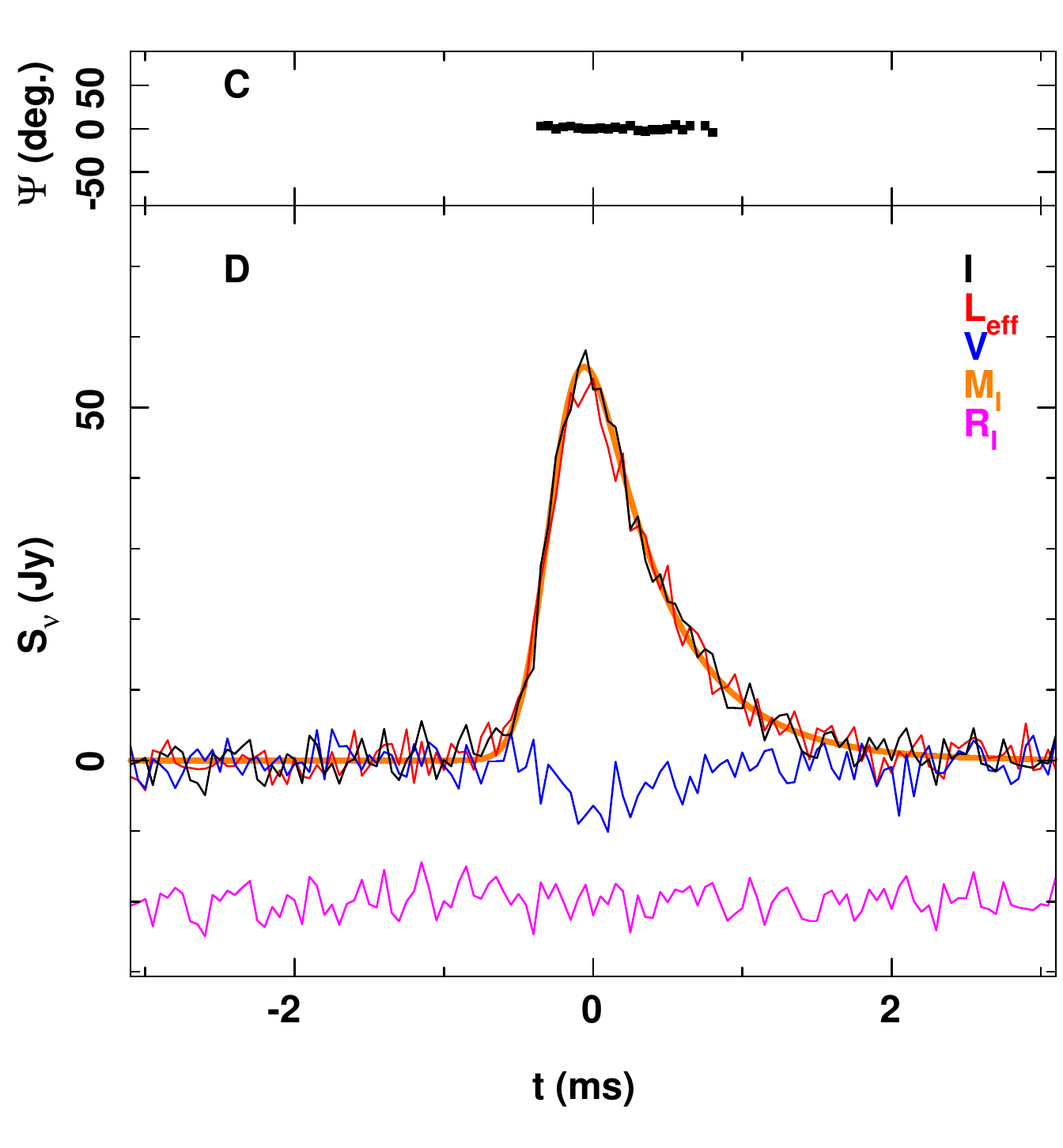}
\end{tabular}

\noindent {\bf Figure 1.} {\bf Radio observations of FRB 20220610A}. (A and B): The burst as observed in the incoherently summed data stream used by the real-time detection system.  The pulse width is dominated by intra-channel dispersion smearing.  Panel~A shows the integrated flux density ($S_\nu$) of the pulse profile as a function of time (t) while Panel~B shows the dedispersed burst dynamic spectrum as a function of frequency (Freq).    The vertical dashed line in panel B indicates the frequency range ($<$1200\,MHz) of high time resolution data that were saved by the pipeline; data at higher frequencies were lost due to latency issues (see text). (C and D): High time resolution data (frequencies $<$1200\,MHz) produced from the raw electromagnetic field samples saved from each telescope after the real-time trigger \cite{online}.
Panel~C shows linear polarization position angle $\Psi$ during the pulse. Panel~D shows the integrated pulse profile in total intensity $I$ (black), linear polarization $L_{\rm eff}$ (red), Stokes $V$ (blue), and a maximum likelihood model of Stokes $I$, $M_I$ (orange). The residuals $R_I$ (pink) between $I$ and $M_I$ are shown offset from zero by $-20$\,Jy for clarity.  The polarisation position angle $\Psi$ is defined such that linear polarization is in one Stokes parameter and hence $\Psi\approx 0$.
\clearpage

\clearpage

\begin{tabular}{c}
    \centering
\hspace{-1in}   
\includegraphics[scale=0.6]{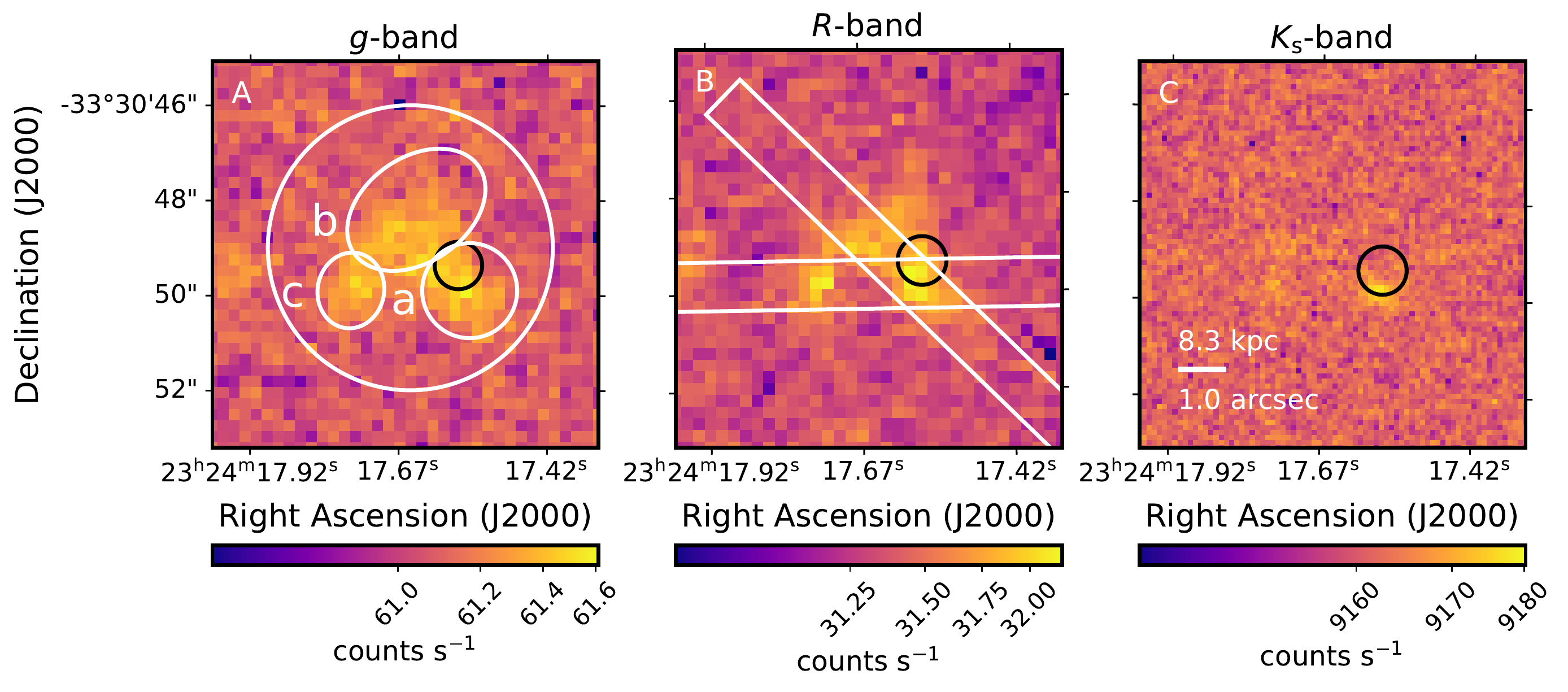}
\end{tabular}
\\
\begin{tabular}{c}
    \centering
   \includegraphics[scale=0.6]{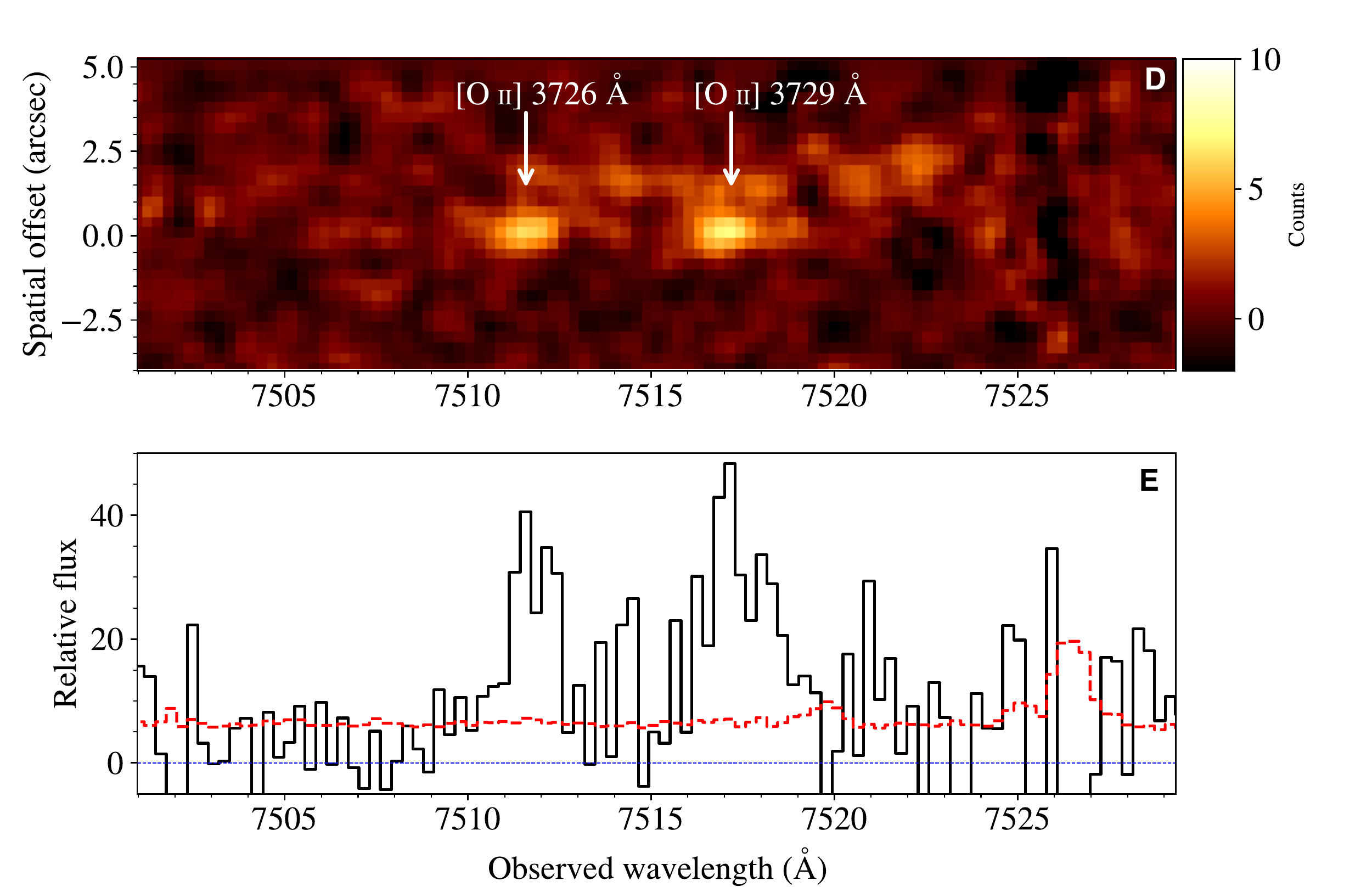}
\end{tabular}

\noindent {\bf Figure 2.    Optical and infrared observations of the host galaxy of \frbname.} (A) $g$-band VLT image.  White ellipses outline the apertures used for the photometry of each component. The larger unlabeled white circle is the aperture used for the entire combined system. The FRB localization and uncertainty (68\% confidence) are indicated by the black ellipse in panels A-C. (B) $R$-band image, with the slit locations used for VLT spectroscopy superimposed in white. (C) $K_{\rm s}$ band image. The same scale bar applies to panels A-C and shows the angular scale and corresponding projected physical scale at the measured host redshift. The color bar below each panel indicates relative count rates in each processed image.
(D) Two-dimensional VLT spectroscopy at a slit position angle on the sky of $45^{\circ}$ covering components (a) and (b) as marked in Panels A and B.
White arrows indicate the two lines due to 
[O\,{\sc ii}] with rest wavelengths of 3726 and 3729~\AA. (E) One-dimensional spectrum (black solid line) and its uncertainty (red dashed line) extracted from the image shown in Panel D, centred on the peak of the [O\,{\sc ii}] lines using an aperture width of 1.8\arcsec. The blue line indicates zero relative flux in arbitrary units. 

\clearpage

\clearpage

\begin{tabular}{c}
\includegraphics[width=12cm]{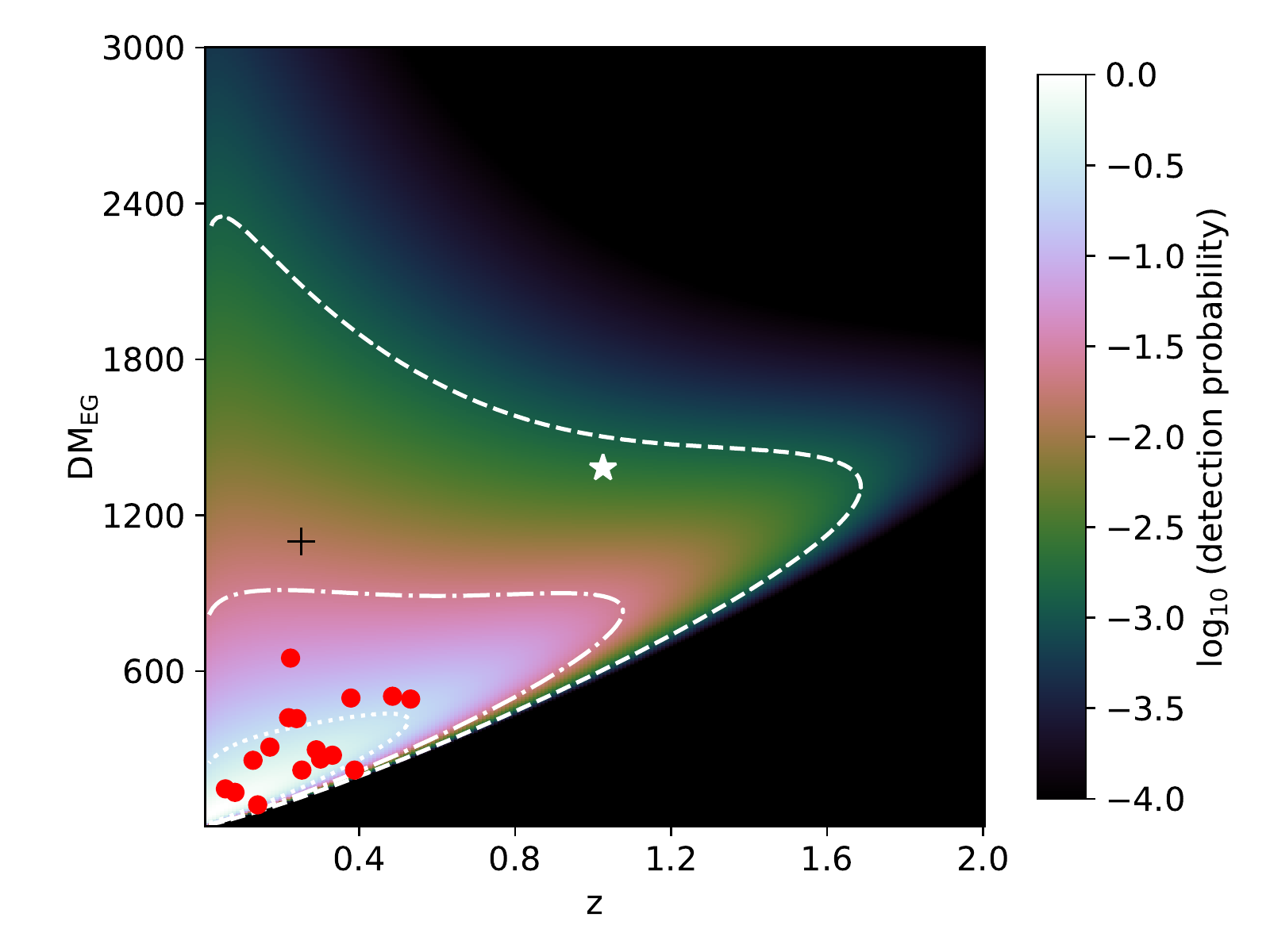} 
\end{tabular}
\\
\noindent {\bf Figure 3.  Relationship between redshift and extragalactic dispersion measure for FRBs.} Data are from the ASKAP incoherent sum survey \cite{2022MNRAS.516.4862J}.  The extragalactic dispersion (DM$_{\rm EG}$) is the contribution to DM after subtraction of a model for the Milky Way \cite{online}. The  localized FRBs are shown as red circles at the host galaxy redshifts.
The color scale indicates an estimated detection probability assuming an increased maximum energy density $\log_{10} E_\nu\,[{\rm erg\,Hz^{-1}}] = 32.7 \pm 0.2$ (found in this work). White contours show 50\% (dotted), 90\% (dash-dot) , and 99\% (dashed) of the probability. \frbname\ is plotted with the white star. The repeating  FRB\,20190520B, shown as the black cross, was not used in the population inference due to varying selection effects between the discovery instruments \cite{James2022Meth}.  The uncertainties for the measurements are smaller than the symbol sizes.

\clearpage

\begin{tabular}{c}
   \includegraphics[width=12cm]{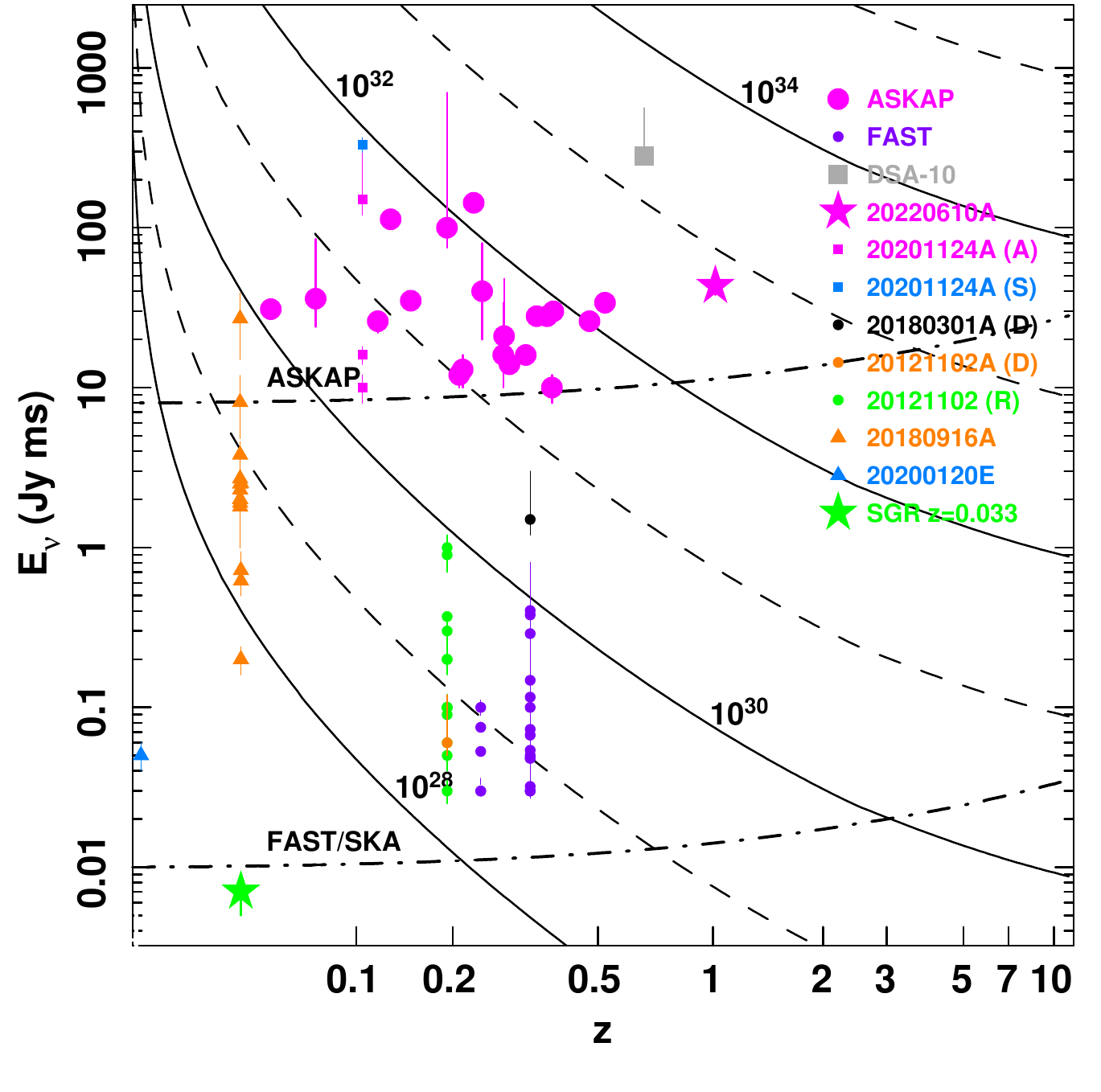} 
\end{tabular}
\\
\noindent {\bf Figure 4.      Logarithmic plot of fluence as a function of redshift for localized FRBs.} The magenta star is \frbname. Other notable FRBs are labeled using their Transient Name Server designations, omitting the FRB prefix for brevity. The curved solid and dashed contours indicate the energy density in units of erg\,Hz$^{-1}$.  The dash-dotted lines show the detection sensitivity of the current ASKAP incoherent sum FRB search system, and Five-hundred-meter Aperture Spherical Telescope and Square Kilometre Array telescopes (labeled FAST/SKA).  Repeating FRBs are shown  at multiple points at the same redshift but different fluences.  Plotting symbols are indicated in the legend. FRB localization instruments are noted in the top right; DSA is the Deep Synoptic Array.
The green star shows the expected fluence of the FRB-like burst from Galactic magnetar soft gamma repeater (SGR) 1935$+$21  \cite{2020Natur.587...54C,2020Natur.587...59B}, if it were placed at the distance of the host galaxy of FRB~20180916A.
The  repeating FRB 20201124A is plotted twice for observations from  ASKAP (labeled ``20201124A (A)'') \cite{2022MNRAS.512.3400K} and the 25-m Stockert radio telescope (labeled ``20201124A (S)'') \cite{2021ATel14556....1H}.  We also show the initial detection of the repeating FRB~20121102A (labeled ``20121102A (D)'') \cite{2014ApJ...790..101S}, and a sample of follow up bursts detected at Arecibo (labeled ``20121102A (R)'') \cite{2019ApJ...876L..23H}; and bursts from two low-redshift repeating FRB sources: FRB~20180916A  \cite{2020Natur.577..190M}  and FRB~20200120E \cite{2023MNRAS.520.2281N}. 
Burst properties are reported from ASKAP \cite{2019Sci...365..565B,20181112,2020Natur.581..391M,2022MNRAS.516.4862J}; DSA-10 \cite{2019Natur.572..352R} and FAST detections of repeating FRB 20180301A \cite{2020Natur.586..693L} and FRB 20190520D  \cite{2022Natur.606..873N}, in addition to the initial detection of FRB 20180301A with the Parkes radio telescope 
(labeled ``20180301A (D)'') \cite{2019MNRAS.486.3636P}.   There are biases in FRB searches which may result in the first detections of repeating sources having different properties than subsequent detections, particularly those made with other facilities \cite{2019ApJ...887L..30K}.

\clearpage

\renewcommand\thesection{S\arabic{section}} 
\setcounter{section}{0} 

\renewcommand\thetable{S\arabic{table}} 
\setcounter{table}{0} 

\renewcommand\thesection{S\arabic{section}} 
\setcounter{section}{0}

\renewcommand\theequation{S\arabic{equation}}
\setcounter{equation}{0}

\renewcommand\thefigure{S\arabic{figure}}
\setcounter{figure}{0}


\begin{center}
{\LARGE
Supplementary materials for \\
``A luminous fast radio burst that probes the Universe at redshift $1$''
}
{\large

 }
\end{center}
\pagenumbering{gobble}

\noindent 
Authors:  S.~D.~Ryder, K.~W.~Bannister, S.~Bhandari, A.~T.~Deller, R.~D.~Ekers, M.~Glowacki, A.~C.~Gordon, K.~Gourdji, C.~W.~James,
C.~D.~Kilpatrick, W.~Lu, L.~Marnoch, V.~A. Moss, J.~X.~Prochaska,
H.~Qiu, E.~M.~Sadler, S.~Simha,  M.~W.~Sammons, D.~R.~Scott, N.~Tejos, \&  R.~M.~Shannon.\\

Correspondence: rshannon@swin.edu.au

\vskip 10mm
\noindent
{\bf This PDF file includes:}\\
Materials and Methods\\
Supplementary Text\\
Fig S1 $-$ S8\\
Tables S1 $-$ S2 \\
References (43 $-$ 108)
\newpage
\pagenumbering{arabic}
    \setcounter{page}{1}

\section{Materials and Methods}

\subsection{Radio observations}

\frbname\ was detected as part of the Commensal Real-time ASKAP Fast Transient (CRAFT) survey \cite{2010PASA...27..272M}, which uses ASKAP\cite{2021PASA...38....9H} to search for dispersed radio transients such as FRBs.  
The PAF receivers on ASKAP enable 36 beams to be digitally synthesized across the focal plane of the antenna. 
The FRB detection system on ASKAP at the time of observation searches intensity data, incoherently summed across telescopes, in real time. Due to constraints on performance of the search system, only a subset of the $36$ antennas were included in the real-time sum, with trade offs between the number of antennas included, and the time resolution of the data searched. During the observation in which \frbname\ was detected, the searches were conducted using $22$ of the antennas, with a time resolution of $1.182$\,ms. In all cases a total bandwidth of $336$\,MHz was searched with a spectral resolution of $1$\,MHz.
The search data streams were recorded to disk in pulsar filterbank\cite{sigproc} format for subsequent analysis whenever a burst was detected.
Details of the search system have been described previously \cite{2019Sci...365..565B}. 

FRB searches with ASKAP are intended to be commensal with other observations. 
However some observations are also conducted in a mode termed filler,  during times where standard observations cannot be scheduled.
 The observations reported here were conducted at a central frequency of $1271.5$\,MHz with the configurable beam in a hexagonal close packed configuration and beam centers separated by $0.9$ degrees.
 FRB\,20220610A was detected during filler observations while pointed towards FRB\,20220501C, a previously-detected FRB \cite{2023arXiv230507022B}. The  different sky locations (the burst positions, with precisions better than one arsecond, are separated by approximately 1.5\,$\deg$) and  burst dispersion measures --- $449.5\pm0.2$\,\pccm\ and \dmfrb, respectively --- rule out the two bursts being from the same repeating source.

The burst was identified with a S/N of 29.8 in the primary beam. The ASKAP beams oversample the focal plane, so the burst also appeared at S/N$>5$ in four adjacent beams.  The burst width, at full width half maximum as measured in the filterbank, was $5.6\pm0.4$\,ms.  This is consistent with the 6.0\,ms burst width expected from intrachannel dispersion smearing \cite{lorimer2005handbook}.

\subsection{FRB localization}

From each ASKAP antenna, 3.1\,s of voltage data was saved after being triggered by the real-time detection system.  Due to the dispersive sweep of 2.04\,s across the 336~MHz-wide ASKAP band, combined with the 2.4\,s latency of the detection system, much of the FRB had already been lost from the voltage buffer before it was frozen and downloaded.  Below a frequency of 1192~MHz, the FRB signal remained present in the saved data. Approximately $2.5\,$h  after the FRB detection, a voltage download was triggered while observing the bright  radio source PKS~0407-658, to be used for calibration purposes, followed shortly after by a voltage download on the Vela pulsar (used as a polarization calibrator).

Our localisation of \frbname\ used the ASKAP FRB astrometric pipeline \cite{2019Sci...365..565B,Day2020}. We correlated a small time window matching the dispersed FRB itself (referred to as the gated dataset), as well as the full 3.1s time window from the FRB voltage download (the field dataset) and the voltages downloaded while observing PKS~0407-658 (the calibrator dataset).  The voltages downloaded while observing the Vela pulsar were also correlated (the polcal dataset).
The downloaded voltage data were processed using the \textsc{celebi} pipeline \cite{2023A&C....4400724S}, which is based on the same procedures as for earlier ASKAP FRBs\cite{2019Sci...365..565B,Day2020}. Voltage data were correlated using the \textsc{difx} software correlator \cite{Deller2011}, to produce the four visibility datasets.

Calibration was derived from the calibrator dataset and applied to the gated, field, and polcal datasets following previously described procedures  \cite{Day2020}.  None of the 22 antennas had to be removed from analysis for reasons such as irregular visibility amplitudes that occasionally affect a subset of downloaded CRAFT data. After calibration, each dataset was imaged using \textsc{casa} \cite{CASA}. Portions of the spectrum affected by radio frequency interference (for the field and polcal datasets) and the lack of FRB emission due to the limited voltage buffer size and FRB dispersion (for the target dataset) were flagged and discarded prior to imaging. The intrinsic width and scattering time of the FRB were fitted using a Bayesian method described below (section \ref{subsec:scatteringFit}). Positions and corresponding uncertainties were measured in the image plane using the \textsc{aips} task \texttt{JMFIT} \cite{2003ASSL..285..109G}.  

The target position was measured from the FRB image to be right ascension $23^{\rm h} 24^{\rm m} 17^{\rm s}.559 \pm 0^{\rm s}.006$, declination $-33^{\circ} 30^{\prime} 49^{\prime\prime}.87 \pm 0^{\prime\prime}.07$, J2000 equinox.  We used an ensemble of seven background radio continuum sources from our 3.1\,s field image to estimate any systematic shift in the FRB position, following published methods \cite{2021PASA...38...50D}.  
These seven sources were selected because they are all the detections in this field image that were unresolved in the National Radio Astronomy Observatory Very Large Array Sky Survey (NVSS) \cite{NVSS}. 
All these sources were also present in the Rapid ASKAP Continuum Survey (RACS) catalog \cite{RACS}.  Neither catalogue is ideal for this comparison; RACS is at a lower frequency than our observations and the catalogue used in this analysis is  affected by astrometric errors of up to $\lesssim$1 arcsecond \cite{RACS} while NVSS is at higher frequency and lower angular resolution.  We measure a small astrometric correction from each catalogue, which are consistent at the 1$\sigma$ level: $-0.16^{\prime\prime} \pm 0.5^{\prime\prime}$ and $0.85^{\prime\prime} \pm 0.5^{\prime\prime}$ in right ascension and declination respectively for the  RACS catalogue positions, and $0.39^{\prime\prime} \pm 0.5 ^{\prime\prime}$ and $0.14 ^{\prime\prime} \pm 0.5 ^{\prime\prime}$ in right ascension and declination respectively from the NVSS catalogue positions.  This offset is small and consistent with previous results \cite{2021PASA...38...50D}.

Because we have no reason to prefer one catalogue over the other, we take an unweighted average of the two offsets as our final correction: $0.12^{\prime\prime}$ and $0.50^{\prime\prime}$ in right ascension and declination respectively, with an uncertainty on the correction of $0.5^{\prime\prime}$ in each coordinate.  This is equivalent to using an unweighted average of the RACS and NVSS reference positions for each source when performing the comparison to the field image positions, without reducing the corresponding reference position uncertainty.  
The uncertainty on our systematic position correction  dominates the final position uncertainty. 
After applying this astrometric correction, the final FRB position is $23^{\rm h} 24^{\rm m} 17^{\rm s}.569 \pm 0^{\rm s}.040$, $-33^{\circ} 30^{\prime} 49^{\prime\prime}.37 \pm 0^{\prime\prime}.50$ (J2000).

\subsection{Beamforming and spectral-temporal polarimetry}

The voltage data were also used to reconstruct a complex-valued time series of the electric field in two linear polarizations, from which we derived the Stokes $I, Q, U, V$ parameters and measurements of the polarization properties of \frbname\ at a time resolution of 3~ns. The beamforming process used to achieve this has been previously outlined  \cite{Cho2020,2023A&C....4400724S}. This pipeline (\textsc{celebi}, \cite{2023A&C....4400724S}) applies per-antenna time delays to account for the difference in signal arrival times and uses a polyphase filterbank inversion to obtain a single complex, high frequency resolution spectrum for each polarization per antenna. Calibration solutions were then applied, before spectra were summed across all antennas and derippled. 
Differential gain and phase between the two linearly polarized feeds was derived using 
the polcal dataset  of the Vela pulsar (PSR~J0835$-$4510). The FRB data were then coherently dedispersed (using an initial DM of \dminit) and a Fourier transform inversion applied. Finally, the Stokes products were calculated and the data averaged in time and frequency to $1\,\mu$s resolution.

To find the S/N-maximising DM, the data were coherently dedispersed to several trial DMs about the nominal value of 1458\,\pccm. For each trial DM, Stokes $I$ was averaged at $10\,\mu$s resolution, and the S/N calculated for sliding boxcar windows of varying width. The peak value of S/N was found at \dmfrb, with the uncertainties --- consistent with our initial estimate of \dminit\ --- corresponding to the range producing a S/N within 1 of the peak. These uncertainties reflect the intrinsic burst structure and not random noise: doubling the amount of noise in the data (by adding random Gaussian deviates with standard deviation equal to that found in off-pulse data), and re-performing the procedure 1000 times, results in 68\% of the data falling in the range of $1458.15_{-0.25}^{+0.05}\,\pccm$.

\subsection{Burst scatter broadening}\label{subsec:scatteringFit}

We use a nested sampling approach  \cite{qiu_population_2020}, implemented using the  \textsc{bilby}\cite{ashton_bilby_2019} Python library to determine the maximum a-posteriori model of the pulse profile. Assuming uniform priors on the pulse broadening time $\tau$ and full width at half maximum $w$, and using a uniform weighting across the burst profile while assuming Gaussian errors to calculate and  maximize the posterior probability of the model, the peak posterior distribution parameters are $\tau=0.511\pm0.012$\,ms and $w=0.407\pm0.012$\,ms.
The \textsc{NE2001} model of the Galactic electron density variations \cite{2002astro.ph..7156C} predicts the Galactic scatter broadening, $\tau_{\rm MW}$, to be 48.3~ns at 1.1475\,GHz, so we infer the scatter broadening is dominated by the host environment, consistent with a large host DM contribution. Fitting our model of pulse broadening to four 22\,MHz sub-bands and assuming a $\tau \propto \nu^{-4}$ scaling yields a consistent result for the scattering time.

Taking the measured redshift of 1.016 along with our reference frequency of $1.1475$\,GHz, and assuming a scattering index of $\tau \propto \nu^{-4}$, we scale to the host galaxy rest frame (indicated with ${}^\prime$) to find $\tau^\prime_{\rm 1.4\,GHz}=1.88 \pm 0.02$\,ms, and $w^\prime = 0.202 \pm 0.006$\,ms. In Table \ref{tab:scattering} we compare our measurements to FRB\,20190520B, another FRB with large excess DM above the Macquart relation, with the same redshift corrections applied.

\subsection{Burst rotation measure}

\begin{figure}[!ht]
    \centering
    \includegraphics[width=12cm]{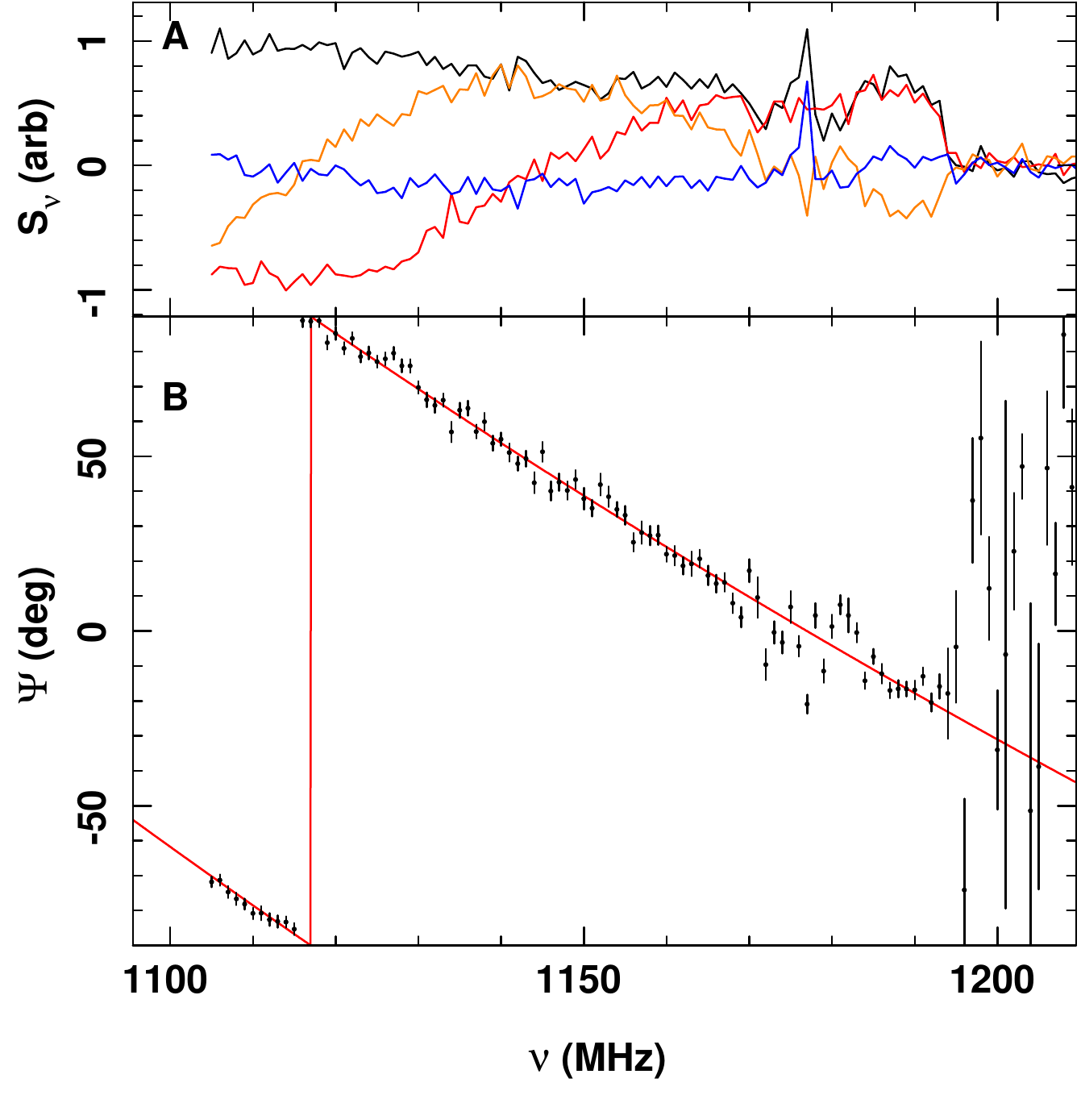}
    \caption{{\bf Pulse-averaged spectrum of FRB 20220610A.} (A) Spectra for total intensity $I$ (black), Stokes $Q$ (red), Stokes $U$ (magenta), and Stokes $V$ (blue). The variation in flux density between $Q$ and $U$ is due to Faraday rotation. The absence of FRB emission above $\sim$1195~MHz is due to buffered voltages above this frequency (see text). (B) Linear polarization position angle $\Psi$ (black points, with 1$\sigma$ error bars). The red line is a maximum-posteriori model, which corresponds to a rotation measure of $215\pm2$~rad~m$^{-2}$.}
    \label{fig:frb_spectrum}
\end{figure}

After polarimetric calibration, we found that \frbname\ was highly linearly polarized, and that the Stokes $Q$ and $U$ parameters showed frequency-dependent variation that we attribute to Faraday rotation in cold plasma.
Because the polarization of the burst is constant in time, we  measure the rotation measure (RM) of the burst after averaging the de-dispersed burst in time to form one-dimensional (time-independent) spectra of the burst Stokes $Q$ and $U$ parameters: $Q(\nu)$ and $U(\nu)$ \cite{2Dfitting}.  We calculated these by integrating over a $1.9$\,ms gate around the peak of the pulse and used a Bayesian inference method described previously \cite{2019Sci...365..565B,2022ascl.soft04008L}. In brief it fits a model of the expected frequency-dependent variations in Stokes $Q$ and $U$, while marginalizing over intensity variations in individual channels. We find the best-fitting  rotation measure to be $215\pm 2$ rad m$^{-2}$.  The pulse-averaged spectrum of the burst and the best-fitting model for the rotation measure-induced change in the polarization position angle $\Psi$ are presented in Figure \ref{fig:frb_spectrum}.

The burst shows strong linear polarization with no evidence for polarization position angle changing across the pulse.  
The noise statistics of this quantity are Gaussian as it is a linear combination of components that also have Gaussian noise distributions, unlike the more general estimate of linear polarization $L = \sqrt{Q^2 +U^2}$  \cite{2001ApJ...553..341E,Day2020}, which requires a debiasing of linear polarization intensity \cite{2001ApJ...553..341E}. Linear and circular polarisation fractions were calculated by integrating (using a box car window) the polarized flux over the on-pulse region, and normalizing by the total intensity.  Uncertainties in linear and circular polarisation were determined by measuring the root mean square of the off-pulse baseline and scaling by the square root of the number of samples averaged in the spectrum.

\subsection{Optical and infrared imaging}

We obtained images of the location of \frbname\ using the FOcal Reducer and low dispersion Spectrograph 2 (FORS2) on Unit Telescope 1 (UT1) of the European Southern Observatory (ESO) Very Large Telescope (VLT) on Cerro Paranal. On-chip binning of 2$\times$2 yielded a pixel scale of 0.25\arcsec\ over the 6.8\arcmin $\times$6.8\arcmin\ field of view. Five exposures of 400~sec in the R\_SPECIAL filter \cite{FORS2}, with offsets of $\pm10$\arcsec\ between each, were obtained on 2022~July~1~UTC, while a total of 10 exposures of 600~sec in the $g$ filter were obtained across 2022 Oct 2-4~UTC, in seeing of $0.9$ to $1.0$\arcsec. The FORS2 images were processed  as described elsewhere \cite{2020Natur.581..391M}, with the photometry calibrated against Dark Energy Camera Local Volume Exploration survey data \cite{DELVE} for the $g$-band images, and using the ESO FORS2 Quality Control level-1 database for $R$-band.

Additional optical imaging of the host of \frbname\ was obtained with the Deep Imaging Multi-Object Spectrograph (DEIMOS) on the Keck-II  telescope on Maunakea, Hawaii on 2022 September 24~UTC in the $V$, $I$, and $z$ filters for 3$\times$350~s in each.  Atmospheric conditions were near-photometric with an average seeing of 0.9\arcsec\ and the field was observed at an average airmass of 1.7.  All imaging was reduced following standard procedures in the Python-based
package {\sc potpyri} \cite{POTPyRI}.
Photometric zero points for the stacked frames were derived using Landolt photometric standards \cite{Landolt92} 
transformed into the DEIMOS photometric bands.

Near-infrared imaging was carried out with the High Acuity Wide-field K-band Imager (HAWK-I) camera \cite{HAWK-I} in combination with the ground-layer adaptive optics system on Unit Telescope 4 of the VLT, covering a 7.5\arcmin$\times$7.5\arcmin\ field at 0.106\arcsec/pixel. Observations in the $K_{\rm s}$ filter were taken on 2022~July~24~UTC, and in the $J$ filter on 2022~September~29~UTC in clear conditions and delivered image quality of $0.4$ to $0.5$\arcsec. For each filter 15~integrations of 10~sec were averaged at each of 16 random offset positions within a 20\arcsec\ box. Image mosaics were processed using \textsc{esoreflex} \cite{ESOReflex}, with photometric calibration undertaken using the Two Micron All-Sky Survey point source catalogue \cite{2MASS}.

For the DEIMOS imaging, aperture photometry (reported in Table~1) was performed using a 3\arcsec\ host radius and a 4\arcsec\ to 7.5\arcsec\ background annulus. For the VLT imaging two approaches were used to obtain the photometry reported in Table 1: first, a single circular aperture with a radius of 3\arcsec\ encompassing the entire source; and second, three ellipses defined by eye (Fig. 2A) from the $R$-band imaging, to individually measure components (a), (b) and (c). The results are listed in  Table~\ref{tab:clump_phot}. Both approaches, including background subtraction, used the \textsc{sep} package \cite{SEP}.  Table \ref{tab:clump_phot} also lists the positions of the clumps as defined by the apertures.

\subsection{Optical and infrared spectroscopy}

Spectroscopy covering the entire ultraviolet through near-infrared wavelength range of 0.3--2.48~$\mu$m was conducted with the X-shooter spectrograph \cite{X-shooter} on Unit Telescope 3 of the VLT. Slit widths of 1.0\arcsec, 0.9\arcsec, and 0.9\arcsec\ yielded spectral resolving powers of 5400, 8900, and 5600 in the ultraviolet/blue (UVB), visible (VIS), and near infrared (NIR) arms, respectively. On-chip binning of 2$\times$2 was employed in the UVB and VIS detectors due to the faintness of the target, without affecting the spectral resolution in these arms.

Observations of the target were made at 2~slit position angles: the first on 2022~July 29~UTC at a position angle of 45$^{\circ}$ covering components (a) and (b); and the second on 2022~September~23/24~UTC at a position angle of 90$^{\circ}$ covering components (a) and (c), as illustrated in Figure~2 (B). At each of these position angles, a total of 7200 and 7548~sec respectively of integration time was accumulated in clear conditions and seeing 0.5 to 1.2\arcsec, with frequent nodding of the target by $\pm$2.5\arcsec\ along the slit to assist with sky subtraction. Observations of the B9~V stars HD~138940 and HD~168852 were obtained to enable removal of telluric features, and of the spectrophotometric standard stars EG~21 and LTT~987 to allow relative flux calibration. 
The spectra were reduced with the \textsc{pypeit} data reduction pipeline \cite{pypeit} using standard techniques for image processing and wavelength calibration. The VIS and NIR science exposures were individually processed and the sky-subtracted
2D spectral images were co-added to maximize S/N.
No emission lines were detected in the UVB arm spectra. 
Figure 2 (D-E) shows the 2D image and 1D extracted spectrum at the slit position angle of $45^{\circ}$.

From the spectra of components (a), (b) and (c) we identified emission lines at 7510 and 7515\,\AA\ as the [O\,{\sc ii}] 3726 and 3729\,\AA\ forbidden transitions of ionized oxygen (Figure 2).
Combining the measurements of these lines we determine the redshift of the host galaxy components (a)--(c) to be \zfrb.


\subsection{Dispersion-measure analysis}
\label{sec:dm_analysis}

Our DM analysis is based on a four-component model \cite{2020Natur.581..391M},
\begin{eqnarray}
{\rm DM} & = & \dmmw\ + \dmhalo\ + \dmcosmic + \frac{\dmhost}{1+z}, \label{eq:dmbudget}
\end{eqnarray}
where $\dmmw$ is the Milky Way disk DM contribution, $\dmhalo$ is the Milky Way halo contribution, $\dmcosmic$ is the cosmic DM contribution and $\dmhost$ is the host DM contribution.

We adopt a contribution from the Milky Way interstellar medium (ISM) of \dmmw = 31 \pccm\ \cite{2002astro.ph..7156C} (cf.\ 13.6 \pccm\ predicted by the YMW model \cite{2017ApJ...835...29Y}), and a halo contribution of \dmhalo\ = 50 \pccm, which is on the low side of predictions \cite{ProchaskaZheng2019}, but consistent with the observation of nearby, low-DM FRBs \cite{CHIME_M81_2021}.
The contribution from the intergalactic medium (IGM), including filaments and intervening halos, is a probabilistic distribution about the mean density of ionized gas in the universe \cite{Inoue2004}. 
We adopt a previously described model \cite{2020Natur.581..391M}, which is implemented in the \textsc{frb} Python library \cite{frb_code}. 
We use cosmology from the results of the {\em Planck} spacecraft \cite{PlanckCosmology2018}, in particular the Hubble constant $\Hnot=67.4$\,\hubbunit, physical baryon density parameter $\omegabhh=0.0224$, and assume a feedback parameter $F=0.32$. 
The parameter $F$ governs the clumpiness of cosmic gas distributions, and hence the variance of \dmcosmic\ about its expectation value.
$F$ is the least well constrained parameter in the model,  with $\log_{10} F > -0.89$ 
at 99.7\% confidence \cite{2023arXiv230507022B}.
At the redshift of \zfrb, this model predicts a mean intrinsic \dmcosmic\ of $805_{-115}^{+270}$\,\pccm\ (1\,$\sigma$ uncertainty). We attribute the remaining contribution to \dmhost, 
which includes the FRB host galaxy halo and ISM, and any plasma in the immediate vicinity of the FRB source. This contribution is suppressed by a factor of $1+z$ (here, $\sim 2$), because the dispersion is induced at higher frequencies than those observed.

Using Eq.~\eqref{eq:dmbudget}, we estimate the probability distribution of \dmhost\ for \frbname, with the results shown in Figure~\ref{fig:dmhost}. 
Because \dmcosmic\ is upwardly skewed, with the most likely DM being below the mean, and a small probability of very large excess DMs due to intersections with cosmic structures, our estimate for \dmhost\, is left-skewed. 
The most likely value in the host rest-frame is $\dmhost = 1304$\,\pccm, with a 68\% confidence interval of $(981,1475)$\,\pccm. However, we cannot exclude the possibility of a negligible host contribution: 7\% of the probability density lies in the unphysical region $\dmhost<0$.

We also calculate the equivalent distribution for FRB~20190520B. This FRB originates from a $z=0.241$ host galaxy \cite{2022Natur.606..873N}. Its observed ${\rm DM}_{\rm 20190520B} = 1204.7$\,\pccm\ implies a host galaxy contribution in the observer frame of $903^{+72}_{-111}$\,\pccm. Using our model for \dmcosmic, \dmhalo, and adopting $\dmmw=60$\,\pccm\ \cite{2022Natur.606..873N}, we find a rest-frame DM of $1172$\,\pccm\ with a 68\% confidence limit of $(1104, 1197)$\,\pccm. The predicted range is smaller due to the lower redshift of FRB~20190520B, and hence lower 
absolute variance in \dmcosmic. This agrees with previous interpretations \cite{Ocker2022b} that the DM of this FRB is dominated by the host galaxy.

Both values are much larger than the median \dmhost\ of 186\,\pccm\ derived from population models of FRBs \cite{2022MNRAS.516.4862J}. The log-normal distribution of \dmhost\ is also shown in Figure~\ref{fig:dmhost} --- while very high values of \dmhost\ are disfavored, they are allowed within the model.

\subsection{Burst energetics}

\frbname\ was measured with a peak S/N of 29.8 (which by definition has an uncertainty of $\pm 1$) in CRAFT ICS data formed from 22 ASKAP antennas. The sensitivity in this configuration scales with the number of antennas, $N_{\rm ant}$, as $\sqrt{ N}_{\rm ant}$ compared to single-antenna searches using ASKAP \cite{2019Sci...365..565B}, which have a noise level of $2.68 \pm 0.02$\,Jy\,ms for a 1\,ms burst \cite{2018Natur.562..386S,2019PASA...36....9J}. We have conducted tests on pulsars by comparing ICS to single-antenna mode, and have confirmed this scaling to within $97.5 \pm 2.5$\%.

The effective width of a burst in ICS data is increased from its incident (i.e.\ intrinsic plus scatter-broadening) width by DM smearing over the 1\,MHz channel bandwidth and by the time resolution of 1.182\,ms. Thus S/N was maximized when the burst was integrated over four time samples. This corresponds to an initial band-averaged fluence estimate of ${\rm S/N} \times N_{\rm ant}^{-0.5} \times (\Delta t)^{0.5} \times {N}_{\rm 1 ms} = 37$\,Jy\,ms. This is a lower limit on the fluence, some of which will lie outside the 4-sample search window used by our real-time detection pipeline. Offline fitting of a Gaussian profile to the data produced a full-width half-maximum of 5.6\,ms and integrated fluence of $43 \pm 5$\,Jy\,ms. Including the potential inefficiency of ICS observations, and correcting for the offset from beam center (sensitivity reduced by 2\% at central frequency), we estimate a band-averaged fluence $F_\nu = 45 \pm 5$\,Jy\,ms.

The isotropic-equivalent energy density of transients is related to the observed fluence via
\begin{equation}
E_\nu =  \frac{4 \pi D_L^2}{(1+z)^2} F_\nu.
\end{equation}
The FRB host galaxy has a redshift of \zfrb, corresponding to a luminosity distance $D_L=6929 \pm 2$\,Mpc using our adopted cosmology \cite{PlanckCosmology2018}. We therefore find an energy density of $E_{\nu} = (6.4\pm 0.7) \times 10^{32}$\,erg\,Hz$^{-1}$
at the mean emission frequency of $2.543$\,GHz.

The spectral behavior of FRBs is complex, with some bursts showing broadband structure, and others being isolated in frequency \cite{2021MNRAS.500.2525K}. On average, bursts appear to be either brighter \cite{2019ApJ...872L..19M} or more common \cite{2022MNRAS.516.4862J,2022arXiv220714316S} at lower frequencies. In the former case, fluence scales as $F_\nu \propto \nu^\alpha$, with $\alpha = -1.5_{-0.3}^{+0.2}$ \cite{2019ApJ...872L..19M}. If so, then a K-correction of $(1+z)^{-\alpha} = 2.85$ should be used to obtain $E_{\nu} = (1.8 \pm 0.2) \times 10^{33}$\,erg\,Hz$^{-1}$ at 1271.5\,MHz;
if not, the K-correction is zero. If the FRB is beamed \cite{2020MNRAS.497.3076C}, or amplified by gravitational lensing \cite{2022MNRAS.517.5216S}, the total energy emitted was less than the isotropic equivalent energy.

\section{Supplementary Text}

\subsection{Nature of the host galaxy}

\begin{figure}
    \centering
\includegraphics[scale=0.6]{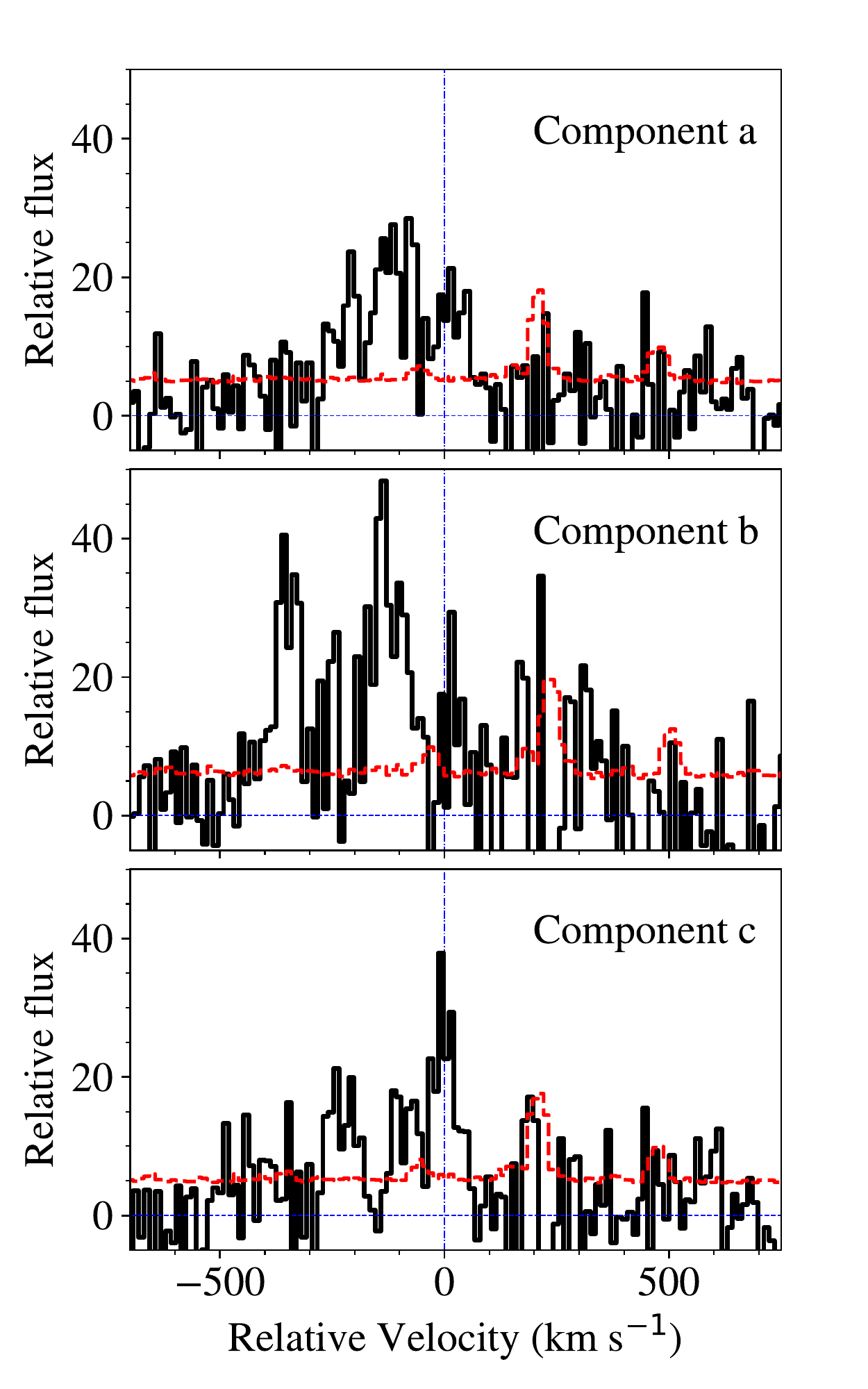}
\caption{ {\bf VLT/X-shooter spectra of the host galaxy of \frbname.} The Figure is the same as Fig 2E, and covers the region of the [O\,{\sc ii}] lines with rest wavelengths 3726 and 3729~\AA, but for the three components  separately and on a velocity scale.    The peak of the [O\,{\sc ii}] 3729~\AA\ line in component (c) is marked as a vertical dashed line. The red dashed line shows the uncertainty in each spectrum.  The blue horizontal line shows zero relative flux on an arbitrary flux scale. }
\label{fig:vlt_spec}
\end{figure}

The spectra in Figure~\ref{fig:vlt_spec} indicate that components (a) and (c) are consistent in redshift to 0.01\%, while component~(b) is blueshifted by a velocity difference of  150~km~s$^{-1}$.
With a projected separation of 16~kpc between compact components (a) and (c), they could be two distinct galaxy nuclei in a merging or interacting system, while component (b) is the gaseous tail linking them and orientated towards the observer.

The measured $(R-K_{\rm s})$ colors for the individual components, as well as for the host system as a whole ($1.70\pm0.18$) are bluer than most  field galaxies at redshifts around 1 \cite{Glazebrook1995, DeBreuck2002, Kajisawa2000}. 
At $z=1$ the $(R-K_{\rm s})$ color approximates rest-frame $(U-J)$ in the AB magnitude system. 
Starburst models \cite{Starburst1999} only produce $(U-J)$ colors close to these values around 10~Myr after the onset of a starburst, or at least 600~Myr later.

To model the stellar population of the host, we use the stellar population synthesis code \texttt{Prospector} \cite{2021ApJS..254...22J} which infers the properties of the stellar population and star formation history based on spectral energy distribution fitting.
We  used the integrated photometry from the entire complex to make an estimate of the host properties as there is insufficient information to characterize individual components.
This effectively assumes that all components have the same star formation history.

We fitted the photometry using the stellar population synthesis library \texttt{python-fsps} \cite{Conroy2009, Conroy2010}.
Additional prior assumptions include a Kroupa initial mass function \cite{Kroupa01}, the Kriek \& Conroy dust-attenuation law \cite{KriekandConroy13}, mass-metallicity relationship \cite{Gallazzi2005}, a ratio between dust attenuation of old versus young stars, and a continuity model non-parametric star formation history (SFH, \cite{Leja2019}) using 8 age bins. 
We used the  \texttt{dynesty} \cite{Speagle2020} package to sample the posterior probability distribution using a dynamic nested sampling algorithm. 
The methodology has been described elsewhere \cite{2023arXiv230205465G}.

\begin{figure} 
    \centering
    \includegraphics[width=10cm]{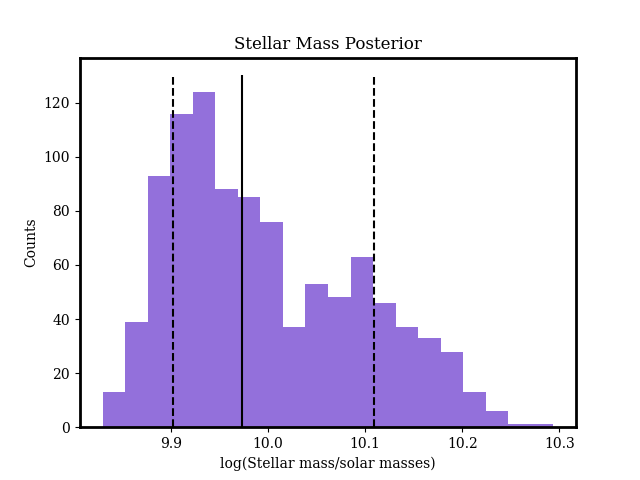}
    \caption{ {\bf Posterior probability distribution of the host galaxy stellar mass as inferred from spectral energy distribution fitting.} The posterior probability has been marginalized over other parameters. The solid line is the median of the distribution while the dashed lines mark the 68\% confidence intervals. The distribution shows bimodality that could be due to the presence of multiple galaxies.}
    \label{fig:prospector_mass}
\end{figure}

\begin{figure} 
    \centering
    \includegraphics[width=10cm]{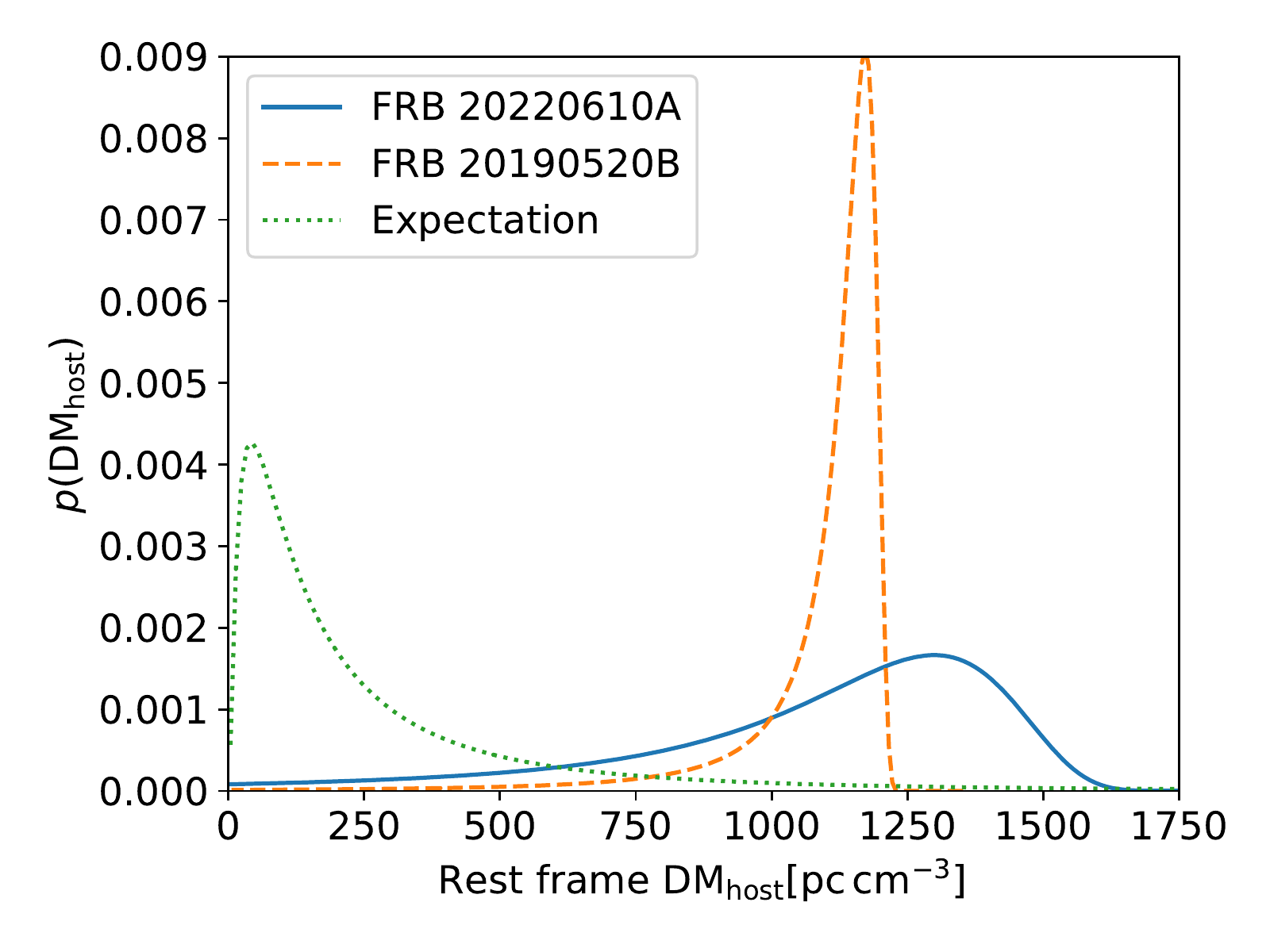}
    \caption{{\bf Probability distributions for the rest-frame host galaxy dispersion measures for two FRBs.}   \frbname\ is shown as a blue solid line and FRB\,20190520B as an orange dashed line. The green dotted line shows the best fitting log-normal distribution for DM$_{\mathrm{host}}$ derived from a population model  \cite{2022MNRAS.516.4862J}.}
    \label{fig:dmhost}
\end{figure}

We find the photometry is most consistent with a galaxy with log(stellar mass/solar masses) of $9.98^{+0.14}_{-0.07} $  at an age of $10.11^{+0.18}_{-0.07}$ Gyr. Additional parameters are reported in Table 1.We show the posterior probability distribution for the stellar mass, marginalized over the other parameters in Figure \ref{fig:prospector_mass}.
The resulting bimodal distribution could be due to the known degeneracy in spectral energy distribution modeling in age, dust, mass, and metallicity \cite{2023arXiv230205465G}. However, given the complex morphology of the host it could instead be due to the presence of multiple stellar populations corresponding to multiple galaxies.  We are unable to distinguish these possibilities.

\subsection{Implications for the maximum FRB energy}

Determining the maximum energy of FRBs, \emax, requires either a sample of localized bursts, or accounting for variations in the Macquart relation. Several parameters --- such as source evolution, spectral  behavior, and the slope of the luminosity function --- are expected to be correlated with \emax, so must be simultaneously fitted when modelling the data. This approach has been taken in previous studies \cite{Luo2020,James2022Lett,2022arXiv220714316S}, which arrived at mutually consistent results. The most precise value reported --- derived from a sample of 16 FRBs with measured redshifts, and 60 without --- is $\lemaxerg=41.26^{+0.27}_{-0.22}$ at 68\% confidence \cite{2022MNRAS.516.4862J}. This calculation assumed an intrinsic emission bandwidth of 1\,GHz, and did not include a K-correction, i.e., it allowed the rate of FRBs to vary with frequency, but not their energy. That study calculated the isotropic equivalent energy, i.e.\ ignored beaming, which is unconstrained by observations. The burst energy function used was an upper incomplete gamma (Schechter) function, with integral slope $\gamma$, and exponential cut-off of the form $\exp(-E/E_{\rm max})$, where $E$ is the bandwidth-integrated burst energy. In this formalism, the burst energy of \frbname\ is $(6.4\pm 0.7) \times 10^{41}$\,erg, 3.5 times the cutoff energy --- which is allowed but unlikely under that model.

\begin{figure}
    \centering
    \includegraphics[width=10cm]{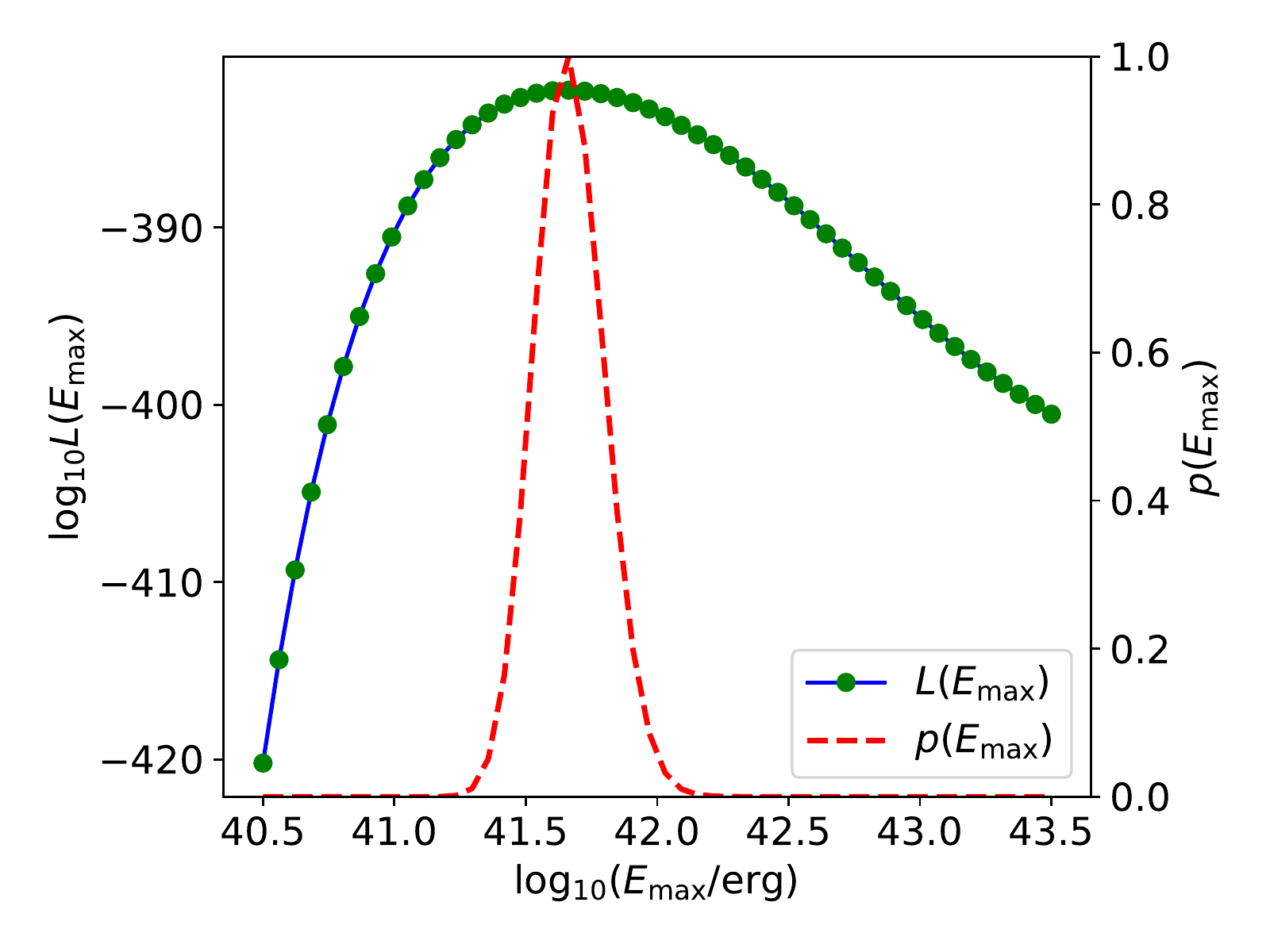}
    \caption{{\bf Estimation of the maximum FRB energy, \emax\ for \frbname.}  Likelihoods $L$ were evaluated after adding \frbname\ to the previous sample of  \cite{2022MNRAS.516.4862J} and varying \emax\ only. The posterior probability distribution $p(\emax)$ assumes a log-uniform prior on \emax.}
    \label{fig:llemax}
\end{figure}

To test the influence of this observation on the derived \emax, we recalculate the model of \cite{2022MNRAS.516.4862J} following a previously described implementation \cite{frb_code,zdm_code}, but add \frbname\ to their sample, and vary \emax\ only. The resulting likelihoods are shown in Figure \ref{fig:llemax}. The peak likelihood is obtained for $\lemaxerg=41.7\pm 0.2$, which is equivalent to $\log_{10} (E_\nu/{\rm erg\,Hz^{-1}}) = 32.7 \pm 0.2$, where the uncertainties were set by assuming a log-uniform prior on \emax\ to produce a Bayesian posterior probability distribution $p(\emax)$. This is 2.7 times higher than the previous estimate, but still within the 90\% confidence interval of the previous model.

Nonetheless, \frbname\ is a high-energy FRB. Given the time and frequency resolutions of the FRB detection algorithm ($t_{\rm samp}=1.182\,$ms and $\Delta \nu = 1$\,MHz respectively), the DM of \dmfrb\ results in instrumental broadening of the signal over a time duration $w_{\rm DM}$ of $5.8$\,ms at band centre.  Including the scattering time $\tau$ and intrinsic width $w$, the total effective width $w_{\rm eff}$ becomes 
\begin{equation}
w_{\rm eff} \approx \sqrt{w_{\rm DM}^2 + t_{\rm samp}^2 + w^2 + \tau^2} = 5.9\,{\rm ms}.
\end{equation}
Without the $w_{\rm DM}$, the width would be 1.35\,ms. Because observed signal-to-noise scales with $w_{\rm eff}^{-0.5}$, the greater width caused by the high DM of the burst decreased our sensitivity by 52\%.

\begin{figure}
    \centering
    \includegraphics[width=10cm]{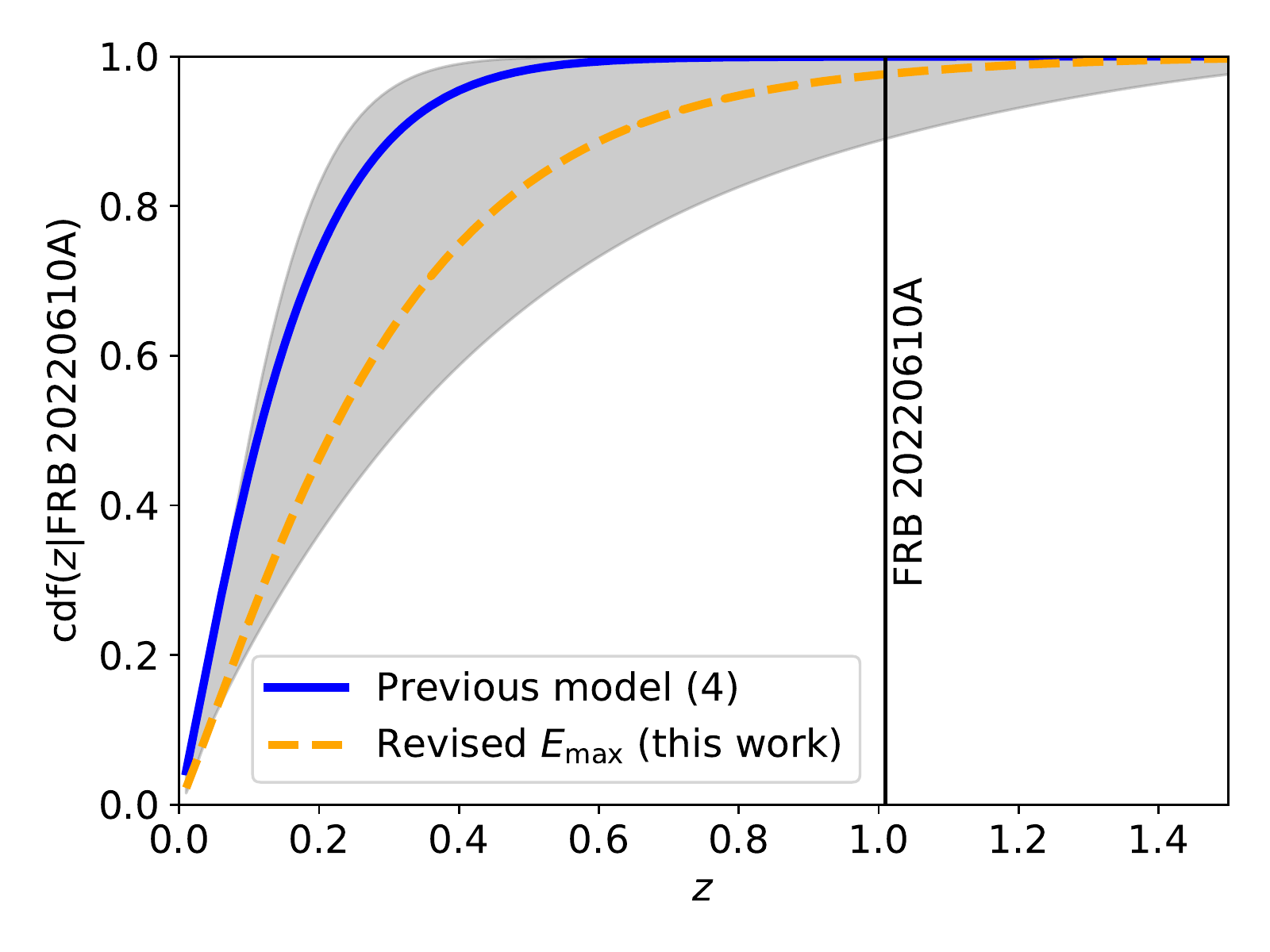}
    \caption{{ \bf Cumulative probability distribution for the redshift, $z$ of \frbname}. This was  estimated using a previous model\cite{2022MNRAS.516.4862J} (blue solid line);  population parameters varied within their 90\% confidence limits (grey shading); and with our revised estimate of \emax\ (orange dashed line).}
    \label{fig:pzgdm}
\end{figure}

Using the observed properties of \frbname\ (DM, width, fluence), the observation parameters at time of detection (location in beam, observation frequency, sampling time, number of antennas), and estimates of the FRB population parameters, we estimate the probability of observing \frbname\ at a given redshift, $p(z|{\rm \frbname})$. The results are plotted in Figure~\ref{fig:pzgdm}. We use the best-fitting population parameters (\emax, $\gamma$, spectral rate scaling index $\alpha$, scaling with the star-formation rate $n_{\rm sfr}$, and log-mean and standard deviation of the host galaxy DM $\mu_{\rm host}$ and $\sigma_{\rm host}$) from the previous model for \Hnot = 67.4\,\hubbunit\ \cite{PlanckCosmology2018}, then set \emax\ to our revised value. We also explore parameter sets corresponding to these same parameters when at their upper and lower 90\% confidence limits. The resulting cumulative probability distributions $cdf(z|{\rm \frbname})$ are plotted in Figure \ref{fig:pzgdm}.

Our model for $cdf(z|{\rm \frbname})$ is based on previous work \cite{2020Natur.581..391M,James2022Meth}, and includes the increased probability of encountering halos with increasing $z$, and the resulting selection effects due to DM smearing. However, the model does not include correlations between DM and the scattering time $\tau$, because there is no observational evidence for such a correlation \cite{James2022Meth}. It has been suggested \cite{2022ApJ...931...88C} that a large observed scattering time $\tau$ correlates with a larger value of $\dmhost$. In such a case, a high value of $\tau$ would be evidence for a higher value of $\dmhost$, hence a lower value of $\dmcosmic$ and thus $z$. We regard this as almost certain, however attempts to quantify it have been highly uncertain, with at least one FRB with a large excess DM having low $\tau$ \cite{Bhandari2022b}. Such a wide variation between $\tau$ and DM is also found for Galactic pulsars \cite{2004ApJ...605..759B}. Therefore, we regard the approach of using $\tau$ to estimate $\dmhost$ as having minimal utility.

\subsection{Limits on depolarization due to multi-path propagation}

The depolarization $f$ due to a spatially variable rotation measure in a foreground scattering screen is \cite{2022Sci...375.1266F,1966MNRAS.133...67B}
\begin{equation}
    f = 1-\exp\left(-2 \lambda^4 \sigma_{\rm RM}^2\right),
\end{equation}
where $f$ is the observed polarization fraction, $\lambda$ is the observing wavelength and $\sigma_{\rm RM}$ parameterizes the RM complexity.
This relation assumes the burst is intrinsically 100\% linearly polarized, and that depolarization occurs in a medium that is in the foreground to the emitting plasma. 

We use our measured lower limit on the depolarization of FRB\,20220610A to set an upper limit on $\sigma_{\rm RM}$ using the relationship 
\begin{equation}
\sigma_{\rm RM} < \sqrt{-\frac{\log(1-f)}{2 \lambda^4}} = 0.57\,{\rm rad}\,{\rm m}^{-2}.
\end{equation}
assuming $f> 0.96$.
In the frame of the host galaxy this corresponds to $\sigma_{\rm RM} < 1.2$ rad\,m$^{-2}$. 

We regard our assumptions about the depolarization as conservative. If the emission is intrinsically less than 100\% linearly polarized we would place a smaller upper limit on $\sigma_{\rm RM}$. While it is theoretically possible that there is depolarization within the emission region, we regard this as unlikely for two reasons.  Firstly, there would need to be non-relativistic plasma within an emission region comprising relativistic electrons \cite{2019ApJ...872L..19M,2022A&ARv..30....2P}.  Secondly there would need to be rotation measure variations in a region of size $ \lesssim c w_{0.5\,{\rm ms}} \Gamma \approx  \Gamma \times 100$\,km, where $c$ is the speed of light, $\Gamma$ is the unknown relativistic beaming fraction and $w_{0.5\,{\rm ms}}$ is the pulse width in units of $0.5$\,ms. This would require a very disordered magnetic field, which we regard as unlikely.

\subsection{Potential origins of the excess dispersion}

 When considering possible origins for the excess dispersion of \frbname, we  make use of  additional information encoded by the intervening plasma on the observed temporal scattering and rotation measure.  In the rest frame of the host galaxy the pulse broadening is 1.88\,ms, and the rotation measure is $860 \pm 8$\,rad\,m$^{-2}$.  Because the burst occurred at high Galactic latitude, the Milky Way contribution to both scattering and rotation measure is small, so we neglect it. 

We first consider the possibility that the excess arises in plasma closely associated with the burst source (i.e., a circumburst medium).
Such an environment has been considered for some repeating FRB sources \cite{2018Natur.553..182M,2022Natur.606..873N}.
The excess dispersion seen in \frbname\ and the repeating FRB\,20190520B are similar, but the pulse broadening is a factor of $6$ higher for FRB\,20190520B, while the rotation measure of FRB\,20190520B is highly variable but reaches values more than an order of magnitude above that of \frbname\ \cite{AnnaThomas2022}. 
The frequency-dependent polarization seen in FRB\,20190520B is consistent with multi-path propagation through a highly magnetized and turbulent local medium that could explain both the scattering and rotation measure; in this model, the rest-frame Faraday dispersion for this source is $\sigma_{\rm RM} =337 $\,rad\,m$^{-2}$, over two orders of magnitude higher than the limit we derive for \frbname\ above. Thus, for the local environment to contribute the bulk of the excess DM seen in \frbname, the medium would need to be less turbulent and less magnetized than for FRB\,20190520B. 

\begin{figure}
    \centering
    \includegraphics[width=10cm]{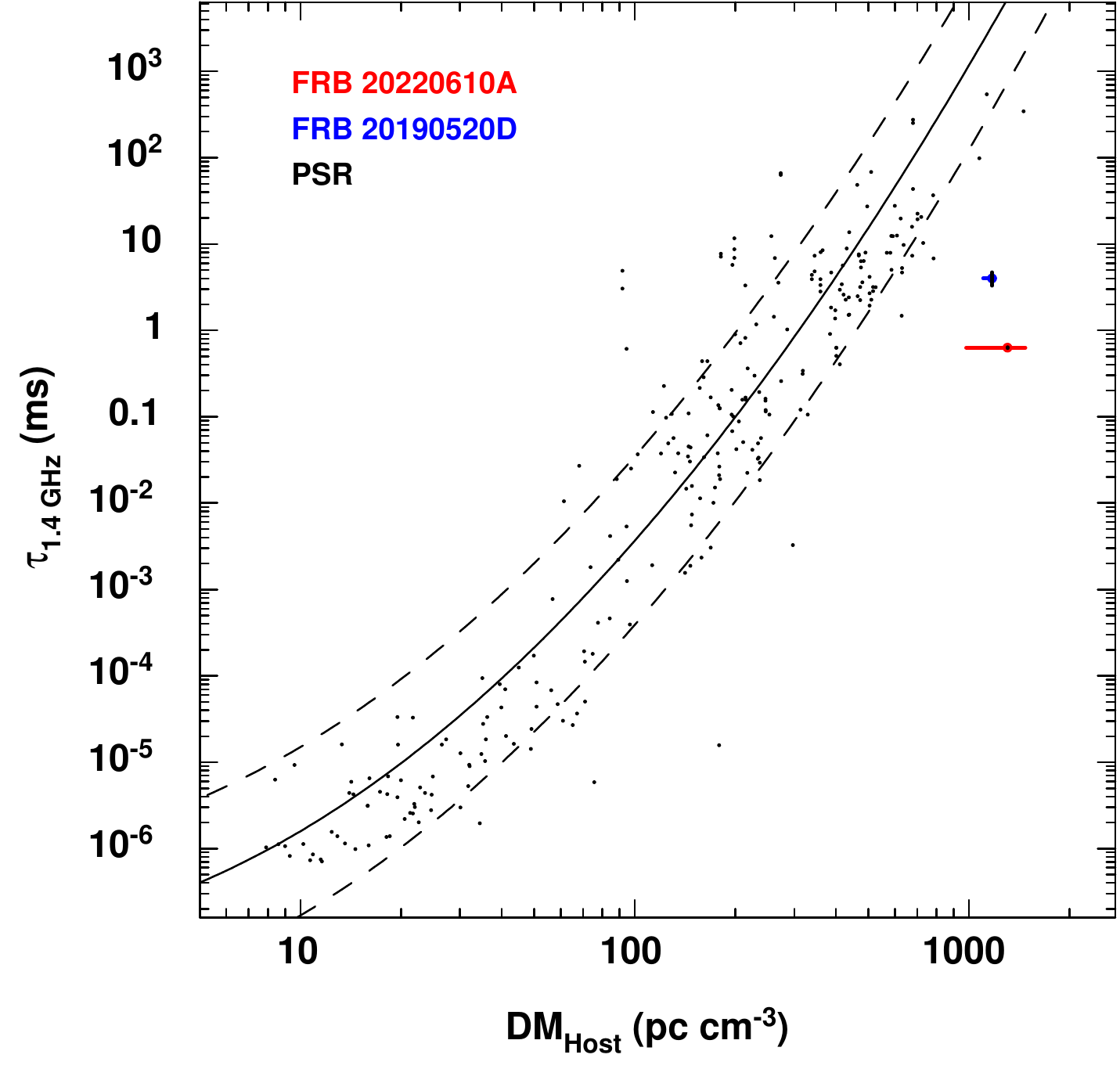}
    \caption{{\bf Dispersion measure and scatter broadening times for Galactic and extragalactic sources.} Milky Way pulsars are labeled PSR (black) \cite{2003astro.ph..1598C,2004ApJ...605..759B}, and FRBs \frbname\ (red) and FRB\,20190520B (blue).  
    The error bars on the FRB DM points reflect uncertainties in the host DM (see text).
    Error bars for pulsar DM are too small to display.
    The solid and dashed lines show a mean and $1\sigma$ relationship derived for pulsars \cite{2004ApJ...605..759B}.    
    } 
    \label{fig:dmtau}
\end{figure}

\begin{figure}
    \centering
    \includegraphics[width=10cm]{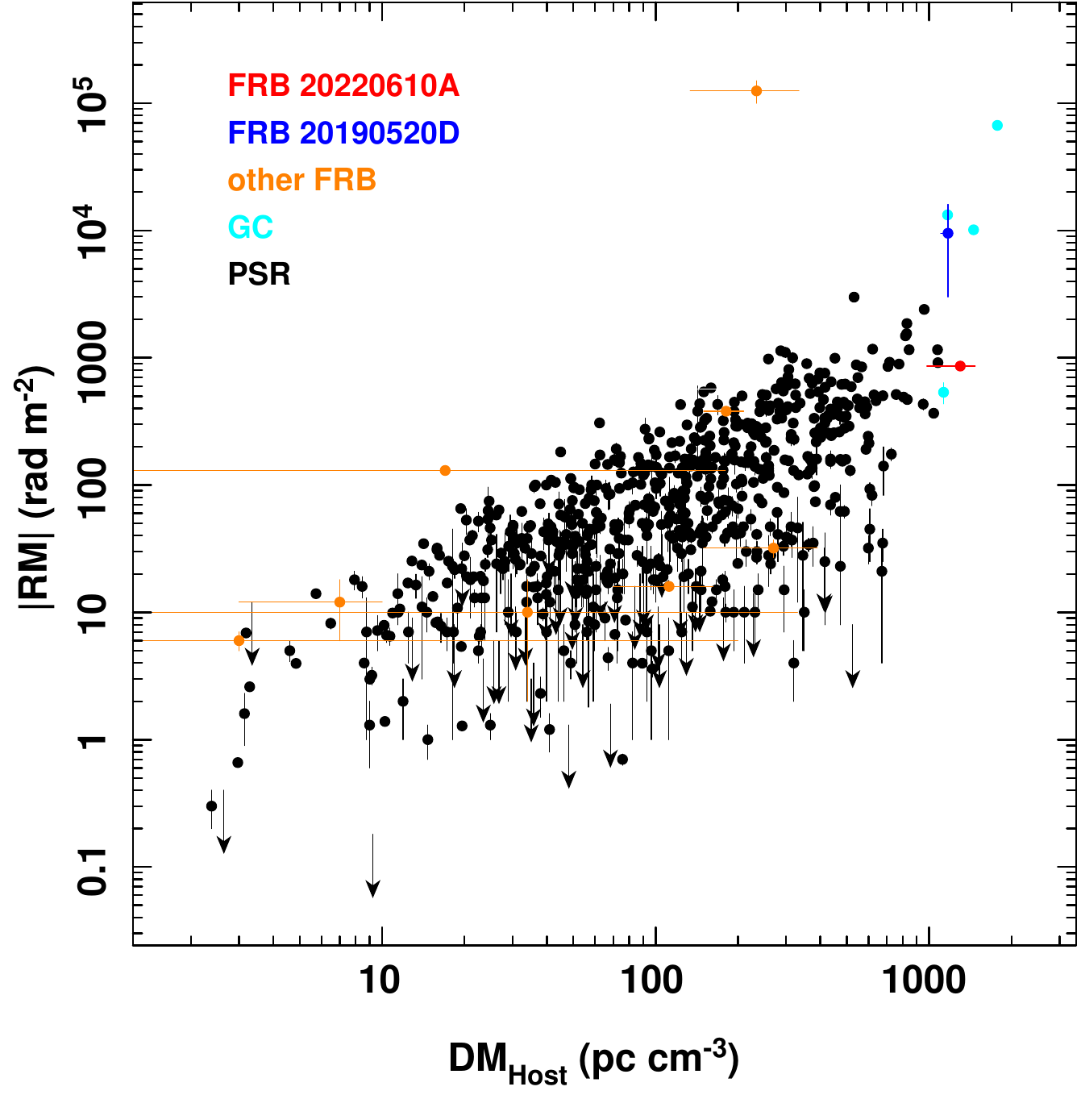}
    \caption{{\bf FRB and pulsar dispersion measures and rotation measures.}   Pulsars \cite{2005AJ....129.1993M} close to the Galactic Center (GC) are highlighted in cyan. Other pulsars (labeled PSR) are in black \cite{2005AJ....129.1993M}, while other FRBs \cite{2022arXiv220915113M} are in orange.    }
    \label{fig:rmdm}
\end{figure}

Another possibility is that this excess arises in the interstellar medium (ISM) of the host galaxy.
We compare the properties of \frbname\ in the host galaxy rest frame to those of FRB\,20190520B and pulsars in the Milky Way. In Figure \ref{fig:dmtau}, we show the distribution of DM and pulse broadening times for pulsars \cite{2004ApJ...605..759B,2002astro.ph..7156C} and these two localized FRBs.
In the case of the FRBs we plot them using the most likely \dmhost\ contribution from our DM analysis, given in Table~\ref{tab:scattering}. 
\frbname\ has substantially lower scattering than expected from a Milky Way-like ISM.
We have decreased the pulse broadening times of the FRBs by a factor of three to account for the plane-wave scattering geometry \cite{2016arXiv160505890C} assumed for FRB pulses scattered in their host galaxy, which contrasts to the spherical wave scattering geometry assumed for pulsar pulses scattered in the Milky Way. 
In Figure \ref{fig:rmdm}, we show the distribution of DM and rotation measures for pulsars \cite{2005AJ....129.1993M} and the two FRBs.  \frbname\ shows a rotation measure similar to most Milky Way pulsars at similar DM, but much lower than the RM associated with pulsars close to the Galactic Centre  \cite{2013MNRAS.435L..29S}, some of which have DMs at the upper end of the allowable range for  \frbname.    

While the Faraday rotation of the burst is not inconsistent with the relationship between dispersion and rotation measure observed in Galactic pulsars, for a Milky Way-like sightline the observed excess DM would imply scatter broadening of $>100$\,ms at 1.4\,GHz (Figure \ref{fig:dmtau}). We therefore disfavor the host galaxy ISM as the sole source of the excess DM for \frbname.

A final possibility is that the excess DM is associated with plasma in an intervening structure along the line of sight, but outside of the host galaxy complex.
However, we do not expect  plasma in the intergalactic medium to have high levels of turbulence or magnetization, so is unlikely to cause the observed rotation measure or pulse broadening \cite{2013ApJ...776..125M,2021ApJ...911..102O}. Therefore we also disfavor this option as the sole source of excess DM for \frbname, unless there is an intervening galaxy halo, where density and turbulence can be higher.

Given these constraints, we cannot  identify a dominant location for the excess DM seen in \frbname, but the observed properties can be plausibly reproduced by a combination of two (or, indeed, all three) of the potential locations.

\clearpage

\clearpage

\clearpage

\clearpage
\newpage

\clearpage

\clearpage

\clearpage

\clearpage

\clearpage

\begin{table}[]
    \centering
        \caption{{\bf Observed scattering properties of two FRBs inferred to have excess DM: \frbname\ and FRB~20190520B \cite{2022Natur.606..873N}}. We scale scattering timescale $\tau$ to 1.4\,GHz in the host galaxy rest frame, assuming $\tau \propto \nu^{-4}$, to calculate $\tau^{\prime}_{\rm 1.4\,GHz}$. The expected contribution from the Milky Way, $\tau_{\rm MW}$, is taken from the NE2001 model \cite{2002astro.ph..7156C}, scaled to $\nu_\tau$. We also include estimates of \dmhost\ from Equation \eqref{eq:dmbudget}, with 68\% confidence intervals.}
    \label{tab:scattering}
    \vskip 5mm
    \begin{tabular}{c c c  }
    \hline \\
         &\frbname\ & FRB\,20190520B \\
        \hline
        $\tau$ (ms) & $0.511\pm0.012$ & $10 \pm 2$ \\
        $\nu_\tau$ (GHz) & 1.1475 & 1.25 \\
        $z$ & \zfrb\ & $0.241 \pm 0.001$ \\
        $\tau^{\prime}_{\rm 1.4\,GHz}$ (ms) & $1.88 \pm 0.02$ & $12 \pm 2$ \\
        $\tau_{\rm MW}$ (ms) &$4.83\times10^{-5}$ & $1.87\times10^{-4}$\\
        \dmhost\ (\pccm)  & $1304_{-323}^{+171}$ & $1172_{-66}^{+27}$\\
        \end{tabular}

\end{table}

\clearpage

\begin{table}[]
    \centering
    \caption{{\bf Photometric measurements, including the positions and sizes of elliptical apertures as shown in Figure~2, for each of the components in the host of \frbname.}}
    \label{tab:clump_phot}
    \vskip 5mm
    \begin{tabular}{c c c c c }
    \hline \\
Component:              &    (a)    &     (b)           &     (c)   & Full aperture\\
        \hline
Right Ascension (J2000) & $23^{\mathrm{h}}24^{\mathrm{m}}17.55^{\mathrm{s}}$ & $23^{\mathrm{h}}24^{\mathrm{m}}17.64^{\mathrm{s}}$ & $23^{\mathrm{h}}24^{\mathrm{m}}17.75^{\mathrm{s}}$ & $23^{\mathrm{h}}24^{\mathrm{m}}17.65^{\mathrm{s}}$
\\
Declination (J2000)     & $-33^\circ30{}^\prime49.9{}^{\prime\prime}$ & $-33^\circ30{}^\prime48.2{}^{\prime\prime}$ & $-33^\circ30{}^\prime49.9{}^{\prime\prime}$ & $-33^\circ30{}^\prime49.0{}^{\prime\prime}$
\\
Position angle (E of N) & $45^{\circ}$   & $-35^{\circ}$  & $-80^{\circ}$   & $0^{\circ}$ \\
Semi-major axis         & $1$\arcsec     & $1.6$\arcsec   & $0.8$\arcsec    & $3$\arcsec\\
Semi-minor axis         & $1$\arcsec     & $1.1$\arcsec   & $0.7$\arcsec    & $3$\arcsec\\
$g$                     & $25.36\pm0.06$ & $25.13\pm0.06$ & $26.01\pm0.08$ & $24.15\pm0.07$\\
$R$                     & $24.98\pm0.05$ & $24.82\pm0.05$ & $25.69\pm0.07$ & $23.78\pm0.06$\\
$J$                     & $23.07\pm0.08$ & $23.40\pm0.13$ & $23.84\pm0.12$ & $21.97\pm0.07$\\
$K_{\rm s}$             & $22.92\pm0.09$ & $>23.66$       & $23.75\pm0.14$ & $22.08\pm0.12$\\
$(R-K_{\rm s})$         & $2.06\pm0.10$  & $<1.16$        & $1.94\pm0.16$  & $1.7\pm0.13$\\
    \end{tabular}
\end{table}


\begin{thebibliography}{100}
\bibitem{2Dfitting}
{When the polarisation position angle varies over the pulse envelope, averaging
  over the pulse profile would cause depolarization and thus reduce the signal
  to noise ratio in polarized intensity} .
\end{thebibliography}
\end{document}